\documentclass[a4paper,prd,notitlepage,aps,11pt,amsmath,amsfonts,amssymb,longbibliography]{revtex4-1}
\usepackage[utf8x]{inputenc}
\usepackage{multirow}
\usepackage[colorlinks,hyperindex,plainpages=false]{hyperref}
\usepackage{graphicx}
\usepackage{mathtools}
\usepackage{upgreek}
\usepackage{esint}

\def\baa#1\eaa{\begin{align*}#1\end{align*}}
\def\bs#1\es{\begin{split}#1\end{split}}
\newcommand{\f}{\frac}
\newcommand{\p}{\partial}
\newcommand{\be}{\begin{eqnarray}}
\newcommand{\ee}{\end{eqnarray}}
\newcommand{\bea}{\begin{eqnarray*}}
	\newcommand{\eea}{\end{eqnarray*}}
\def\ba#1\ea{\begin{align}#1\end{align}}
\def\bsub#1\esub{\begin{subequations}#1\end{subequations}}

\def \bit{\begin{itemize}\setlength\itemsep{0em}} 
\def \eit{\end{itemize}}

\def \k{\kappa}
\def \l{\left}
\def \r{\right}

\def \A{\mathcal{A}}
\def \B{\mathcal{B}}
\def \F{\mathcal{F}}
\def \I{\mathcal{I}}

\def \U{\mathcal{U}}

\def \m{\mu}
\def \n{\nu}

\def \cov{\nabla}
\def \P{\Phi}

\def \a{\alpha}
\def \b{\beta}
\def \g{\gamma}

\def \ra{\rightarrow}

\def \barP{\bar{\Phi}}

\def \barF{\bar{\mathcal{F}}}

\newcommand{\order}[2]{\overset{\mathclap{\scriptstyle(#2)}}{#1}\vphantom{#1}}

\begin{document}

\date{\today}

\title{Post-Newtonian parameters $\gamma$ and $\beta$ of scalar-tensor gravity for a homogeneous gravitating sphere}

\author{Manuel Hohmann}
\email{manuel.hohmann@ut.ee}
\affiliation{Laboratory of Theoretical Physics, Institute of Physics, University of Tartu, W. Ostwaldi 1, 50411 Tartu, Estonia}

\author{Andreas Sch\"arer}
\email{andreas.schaerer@physik.uzh.ch}
\affiliation{Department of Physics, University of Zurich,
Winterthurerstrasse 190, 8057 Zurich, Switzerland}

\begin{abstract}
We calculate the parameters $\gamma$ and $\beta$ in the parametrized post-Newtonian (PPN) formalism for scalar-tensor gravity (STG) with an arbitrary potential, under the assumption that the source matter is given by a non-rotating sphere of constant density, pressure and internal energy. For our calculation we write the STG field equations in a form which is manifestly invariant under conformal transformations of the metric and redefinitions of the scalar field. This easily shows that also the obtained PPN parameters are invariant under such transformations. Our result is consistent with the expectation that STG is a fully conservative theory, i.e., only $\gamma$ and $\beta$ differ from their general relativity values $\gamma = \beta = 1$, which indicates the absence of preferred frame and preferred location effects. We find that the values of the PPN parameters depend on both the radius of the gravitating mass source and the distance between the source and the observer. Most interestingly, we find that also at large distances from the source $\beta$ does not approach $\beta = 1$, but receives corrections due to a modified gravitational self-energy of the source. Finally, we compare our result to a number of measurements of $\gamma$ and $\beta$ in the Solar System. We find that in particular measurements of $\beta$ improve the previously obtained bounds on the theory parameters, due to the aforementioned long-distance corrections.







\end{abstract}

\maketitle

\section{Introduction}


General relativity (GR) is currently the most established theory of gravitation. It correctly describes a number of observations, such as planetary orbits in the Solar System, the motion of masses in the Earth's gravitational field \cite{BETA.Will:2014kxa}, the recently discovered gravitational waves \cite{BETA.Ligo} or the $\Lambda$CDM model in cosmology \cite{BETA.Planck}. However successful on these scales, GR itself does not provide sufficient answers to fundamental open questions such as the reason for the accelerated expansion of the universe, the phase of inflation or the nature of dark matter. Further tension arises from the fact that so far no attempt to extend GR to a full quantum theory has succeeded.

GR is expected to be challenged by different upcoming experiments on ground and in space, such as high precision clocks \cite{BETA.cacciapuoti2011atomic} and atom interferometers in Earth orbits, pulsar timing experiments \cite{BETA.Pulsar} and direct observations of black hole shadows \cite{BETA.Goddi,BETA.Broderick}. This plethora of existing and expected experimental data, together with the tension with cosmological observations, motivates studying alternative theories of gravitation \cite{BETA.Nojiri}. In particular, the upcoming experiments are expected to give more stringent constraints on the parameter spaces of such theories or even find violations of GR's predictions.

One class of alternative theories are scalar-tensor theories of gravity (STG) - an extension to GR that contains a scalar degree of freedom in addition to the metric tensor.
The detection of the Higgs proved that scalar particles exist in nature \cite{BETA.higgs} and scalar fields are a popular explanation for inflation \cite{BETA.inflation.guth} and dark energy \cite{BETA.Quintessence.and.the.Rest.of.the.World}.
Further, effective scalar fields can arise, e.g., from compactified extra dimensions \cite{BETA.compactified.extra.dimensions} or string theory \cite{BETA.Damour}.

While the motivation for such alternative theories of gravitation is often related to cosmology, of course any such theory must also pass Solar System tests. The most prominent class of such tests is based on the post-Newtonian limit of the theory under consideration, which is usually discussed in the parametrized post-Newtonian (PPN) framework \cite{BETA.will.book,BETA.Will:2014kxa}.
It allows characterizing theories of gravitation in the weak field limit in terms of a number of parameters, that can be calculated from the field equations of the theory, and will, in general, deviate from the parameters predicted by general relativity. These parameters can be constrained using observational data and experiments \cite{BETA.Fomalont:2009zg,BETA.Bertotti:2003rm,BETA.Hofmann:2010,BETA.Verma:2013ata,BETA.Devi:2011zz}.
In this work we are interested in the parameters $\gamma$ and $\beta$ only, as these are the only parameters that may differ in fully conservative gravity theories, to which also STG belongs~\cite{BETA.will.book}.

The most thoroughly studied standard example of a scalar-tensor theory is Brans-Dicke theory \cite{BETA.Brans-Dicke.1961}, which contains a massless scalar field, whose non-minimal coupling to gravity is determined by a single parameter $\omega$. This theory predicts the PPN Parameter $\gamma = (1+\omega)/(2+\omega)$, in contrast to $\gamma=1$ in GR. Both theories predict $\beta = 1$. Adding a scalar potential gives the scalar field a mass, which means that its linearized field equation assumes the form of a Klein-Gordon equation, which is solved by a Yukawa potential $\sim e^{-m r}/r$ in the case of a point-like source. In this massive scalar field case, the PPN parameter $\gamma$ becomes a function of the radial coordinate $r$ \cite{BETA.Olmo1,*BETA.Olmo2,BETA.Perivolaropoulos}.

Scalar-tensor theories can be expressed in different but equivalent conformal frames. This means that the form of the general scalar-tensor action is invariant under conformal transformations of the metric, which depend on the value of the scalar field. There are two such frames that are most often considered:
In the Jordan frame, test particles move along geodesics of the frame metric while in the Einstein frame, the scalar field is minimally coupled to curvature.
The PPN parameters $\gamma$ and $\beta$ for scalar-tensor theories with a non-constant coupling have been calculated in the Jordan \cite{BETA.HohmannPPN2013,*BETA.HohmannPPN2013E} and in the Einstein frame \cite{BETA.SchaererPPN2014}.
These works consider a spacetime consisting of a point source surrounded by vacuum.
As will be elucidated below, this assumption leads to problems when it comes to the definition and calculation of the PPN parameter $\beta$.

Applying conformal transformations and scalar field redefinitions allows to transform STG actions, field equations and observable quantities between different frames. It is important to note that these different frames are physically equivalent, as they yield the same observable quantities~\cite{BETA.Postma:2014vaa,BETA.Flanagan}. Hence, STG actions which differ only by conformal transformations and field redefinitions should be regarded not as different theories, but as different descriptions of the same underlying theory.
This observation motivates the definition of quantities which are invariant under the aforementioned transformations, and to express observable quantities such as the PPN parameters or the slow roll parameters characteristic for models of inflation fully in terms of these invariants~\cite{BETA.JarvInvariants2015,BETA.KuuskInvariantsMSTG2016,BETA.Jarv:2016sow,BETA.Karam:2017zno}.

The PPN parameters $\gamma$ and $\beta$ were calculated for a point source \cite{BETA.HohmannPPN2013,*BETA.HohmannPPN2013E,BETA.SchaererPPN2014}, and later expressed in terms of invariants~\cite{BETA.JarvInvariants2015,BETA.KuuskInvariantsMSTG2016}. However, the assumption of a point source leads to a number of conceptual problems. The most important of these problems is the fact that, in terms of post-Newtonian potentials, the Newtonian gravitational potential becomes infinite at the location of the source, so that its gravitational self-energy diverges. It is therefore impossible to account for possible observable effects caused by a modified gravitational self-energy of the source in a theory that differs from GR. We therefore conclude that the assumption of a point source is not appropriate for a full application of the PPN formalism to STG. This has been realized earlier in the particular case of STG with screening mechanisms~\cite{BETA.SchaererPPN2014,BETA.Zhang:2016njn}.

The goal of this article is to improve on the previously obtained results for the PPN parameters $\gamma$ and $\beta$ for a general class of scalar-tensor theories, in which the divergent gravitational self-energy has been neglected. Instead of a point mass source, the gravitating mass source we consider in this article is given by a sphere with homogeneous density, pressure and internal energy that is surrounded by vacuum. In this case the gravitational self-energy remains finite, and can therefore be taken into account. During our calculation we do not fix a particular frame, but instead make use of the formalism of invariants mentioned above already from the beginning in order to calculate the effective gravitational constant as well as the PPN parameters $\gamma$ and $\beta$.

The article is structured as follows. In Sec. \ref{sec:theory} we discuss the scalar-tensor theory action, the field equations and the invariants.
The perturbative expansion of relevant terms is outlined in Sec. \ref{sec PPN Expansion} and the expanded field equations are provided in Sec. \ref{sec Expanded field equations}.
Next, in Sec. \ref{sec Massive field and spherical source}, these are solved explicitly for a non-rotating homogeneous sphere and the PPN parameters are derived.
Sec. \ref{sec Comparison to observations} applies our results to observations.
Finally, we conclude with a discussion and outlook in Sec.~\ref{sec Conclusion}.
The main part of our article is supplemented by Appendix~\ref{app coefficients}, in which we list the coefficients appearing in the post-Newtonian field equations and their solutions.


\section{Theory}\label{sec:theory}
We start our discussion with a brief review of the class of scalar-tensor tensor theories we consider. The most general form of the action, in a general frame, is displayed in section~\ref{ssec:action}. We then show the metric and scalar field equations derived from this action in section~\ref{ssec:feqns}. Finally, we provide the definition of the relevant invariant quantities, and express the field equations in terms of these, in section~\ref{ssec:invariants}.

\subsection{Action}\label{ssec:action}

We consider the class of scalar-tensor gravity theories with a single scalar field \(\Phi\) besides the metric tensor \(g_{\mu\nu}\), and no derivative couplings. Its action in a general conformal frame is given by~\cite{BETA.Flanagan}
\ba
\label{BETA.equ: action}
S = \f{1}{2\k^2} \int d^4x \sqrt{-g}
\l\{ \A(\P) R - \B(\P) g^{\m\n} \p_\m \P \p_\n \P
- 2 \k^2 \U(\P)\r\}
+ S_m [e^{2\a(\P)} g_{\m\n},\chi] \,.
\ea
Any particular theory in this class is determined by a choice of the four free functions $\A, \B, \U$ and $\a$, each of which depends on the scalar field $\P$.
The function $\B$ determines the kinetic energy part of the action. The scalar potential is given by $\U$; a non-vanishing potential may be used to model inflation, a cosmological constant or give a mass to the scalar field.
The last part \(S_m\) is the matter part of the action. The matter fields, which we collectively denote by $\chi$, couple to the so-called Jordan frame metric $e^{2\a(\P)} g_{\m\n}$. It is conformally related to the general frame metric $g_{\m\n}$. The latter is used to raise and lower indices and determines the spacetime geometry in terms of its Christoffel symbols, Riemann tensor and further derived quantities.
In general, the scalar field is non-minimally coupled to curvature. This coupling is determined by the function $\A(\P)$.

There are different common choices of the conformal frame; see~\cite{BETA.JarvInvariants2015} for an overview. In the Jordan frame, one has \(\a = 0\) and the matter fields couple directly to the metric \(g_{\m\n}\). By a redefinition of the scalar field one may further set $\A \equiv \P$. Typically, one considers the coupling function $\omega(\Phi) \equiv \B(\P) \P$. This particular choice of the parametrization is also known as Brans-Dicke-Bergmann-Wagoner parametrization.

Another possible choice for the conformal frame is the Einstein frame, in which the field couples minimally to curvature, $\A \equiv 1$. However, in this case the matter fields in general do not couple to the frame metric directly, but through a non-vanishing coupling function $\alpha \neq 0$. In this case one may also choose the canonical parametrization $B \equiv 2$.

We call the scalar field minimally coupled if the Jordan and Einstein frames coincide, i.e., if one can achieve $\A \equiv 1$ and $\a \equiv 0$ simultaneously through a conformal transformation of the metric.

\subsection{Field equations}\label{ssec:feqns}

The metric field equations are obtained by varying the action \eqref{BETA.equ: action} with respect to the metric. Written in the trace-reversed form they are
\ba \bs
\label{BETA.equ: tensor field equation trace reversed long}
R_{\m\n}
&- \frac{\A'}{\A} \l( \nabla_\m \cov_\n \P + \f12 g_{\mu\nu} \square \P \r)
- \l( \frac{\A''}{\A} + 2\F - \frac{3 {\A'}^2 }{2 \A^2} \r) \p_\m \P \p_\n \P
\\
&- \f12 g_{\mu\nu} \frac{\A''}{\A} g^{\rho\sigma} \p_\rho \P \p_\sigma \P
- \frac{1}{\A} g_{\m\n} \k^2 \U
= \frac{\k^2}{\A} \l( T_{\mu\nu} - \f12 g_{\m\n} T \r) \,,
\es \ea
where we use the d'Alembertian $\square X \equiv \cov^2 X = g^{\m\n} \cov_\mu \cov_\nu X$ and the notation $X' \equiv \f{\p X}{\p\P}$.
Taking the variation with respect to the scalar field gives the scalar field equation
\ba \bs
\label{BETA.equ: scalar field equation}
\F \, \square \P
&+ \f12 \l( \F' + 2 \F \f{\A'}{\A} \r) g^{\m\n} \p_\m \P \p_\n \P
+ \f{\A'}{\A^2} \k^2 \U
- \f{1}{2 \A} \k^2 \U'
= \k^2 \f{\A' - 2 \A \a' }{4 \A^2} T \,.
\es \ea
The function $\F$ introduced on the left hand side is defined by
\ba
\label{BETA.F}
\F \equiv \frac{2 \A \B + 3 {\A'}^2}{4 \A^2} \,.
\ea
Note that these equations simplify significantly in the Einstein frame $\A \equiv 1$ and $\a \equiv 0$. We will make use of this fact in the following, when we express the field equations in terms of invariant quantities.

Further, note that the functions $\A$ and $\B$ should be chosen such that $\F > 0$. A negative $\F$ would lead to a positive kinetic term in the Einstein frame, causing a ghost scalar field that should be avoided. 


\subsection{Invariants}\label{ssec:invariants}

Given a scalar-tensor theory in a particular frame, it can equivalently be expressed in a different frame by applying a Weyl transformation of the metric tensor $g_{\m\n} \ra \bar{g}_{\m\n}$
and a reparametrization of the scalar field $\P \ra \bar{\P}$
\bsub
\label{BETA.equ: transformations}
\ba
\label{BETA.equ: Weyl reparametrization}
g_{\m\n} &= e^{2\bar{\g}(\bar{\P})} \bar{g}_{\m\n} \,,
\\
\label{BETA.equ: scalar field redefinition}
\P &= \bar{f} (\bar{\P}) \,.
\ea
\esub

We defined $\F$ in \eqref{BETA.F} since it transforms as a tensor under scalar field redefinition and is invariant under Weyl transformation,
\ba \bs
\F &= \l( \f{\p \barP}{\p \P} \r)^2 \barF \,.
\es \ea

In order to have a frame independent description, we want to express everything in terms of invariants, i.e., quantities that are invariant under the transformations given above. The matter coupling and the scalar potential can be written in an invariant form by introducing the two invariants~\cite{BETA.JarvInvariants2015}
\bsub
\ba
\mathcal{I}_1(\Phi) = \frac{e^{2\alpha(\Phi)}}{\mathcal{A}(\Phi)} \,,
\\
\mathcal{I}_2(\Phi) = \frac{\mathcal{U}(\Phi)}{\mathcal{A}^2(\Phi)} \,.
\ea
\esub
Given the action in a general frame, we can define the invariant Einstein and Jordan frame metrics by
\bsub
\label{BETA equ: Einstein and Jordan frame metric}
\ba
\label{BETA equ: Einstein frame metric}
g^{\mathfrak{E}}_{\mu\nu} := \mathcal{A}(\Phi) g_{\mu\nu} \,,
\\
\label{BETA equ: Jordan frame metric}
g^{\mathfrak{J}}_{\mu\nu} := e^{2\alpha(\Phi)} g_{\mu\nu}\,,
\ea
\esub
which are related by
\ba
\label{BETA equ: Einstein Jordan frame metric relation}
g^{\mathfrak{J}}_{\mu\nu} = \I_1 g^{\mathfrak{E}}_{\mu\nu} \,.
\ea
Note that if the action is already given in the Einstein frame, the metric coincides with the Einstein frame metric defined above, $g_{\m\n} = g^{\mathfrak{E}}_{\mu\nu}$, and the same holds for the Jordan frame.
We define the Einstein frame metric \eqref{BETA equ: Einstein frame metric} as it significantly simplifies the field equations.
The metric field equations reduce to
\ba
\label{equ: full metric field equation E-frame}
R^{\mathfrak{E}}_{\mu\nu} - 2 \F \, \p_{\m}\P \p_{\n} \P - \kappa^{2}g^{\mathfrak{E}}_{\mu\nu}\mathcal{I}_2 = \kappa^2 \bar{T}^{\mathfrak{E}}_{\mu\nu}\,,
\ea
where
\ba
\bar{T}^{\mathfrak{E}}_{\mu\nu} = T^{\mathfrak{E}}_{\mu\nu} - \frac{1}{2}g^{\mathfrak{E}}_{\mu\nu}T^{\mathfrak{E}}\,, \quad T^{\mathfrak{E}} = g^{\mathfrak{E}\,\mu\nu}T^{\mathfrak{E}}_{\mu\nu} = \frac{T}{\mathcal{A}^2}\,, \quad T^{\mathfrak{E}}_{\mu\nu} = \frac{T_{\mu\nu}}{\mathcal{A}}\,.
\ea
is the trace-reversed energy-momentum tensor in the Einstein frame. It is invariant under conformal transformations and field redefinitions, since also the left hand side of the field equations~\eqref{equ: full metric field equation E-frame} is invariant. Note that we use the invariant Einstein metric \(g^{\mathfrak{E}}_{\mu\nu}\) for taking the trace and moving indices here, in order to retain the invariance of this tensor. For later use, we also define the invariant Jordan frame energy-momentum tensor
\ba
\bar{T}^{\mathfrak{J}}_{\mu\nu} = T^{\mathfrak{J}}_{\mu\nu} - \frac{1}{2}g^{\mathfrak{J}}_{\mu\nu}T^{\mathfrak{J}}\,, \quad T^{\mathfrak{J}} = g^{\mathfrak{J}\,\mu\nu}T^{\mathfrak{J}}_{\mu\nu} = \frac{T}{e^{4\alpha(\Phi)}}\,, \quad T^{\mathfrak{J}}_{\mu\nu} = \frac{T_{\mu\nu}}{e^{2\alpha(\Phi)}}\,.
\ea
Similarly to the metric field equations, we obtain the scalar field equation \eqref{BETA.equ: scalar field equation}
\ba
\label{equ: full scalar field equation}
\F g^{\mathfrak{E}\,\mu\nu} \p_{\m}\p_{\nu}\Phi
- \F g^{\mathfrak{E}\,\mu\nu} \Gamma^{\mathfrak{E}\,\rho}{}_{\n\m}\p_{\rho}\P
+ \f{\F'}{2} g^{\mathfrak{E}\,\mu\nu} \partial_{\mu}\Phi \partial_{\nu}\Phi
- \frac{\kappa^2}{2}{\mathcal{I}_{2}}'
= -\f14 \kappa^2 {(\ln\mathcal{I}_1)}' T^{\mathfrak{E}}\,.
\ea
These are the field equations we will be working with. In order to solve them in a post-Newtonian approximation, we will perform a perturbative expansion of the dynamical fields around a flat background solution. This will be done in the following section.


\section{PPN formalism and expansion of terms}
\label{sec PPN Expansion}
In the preceding section we have expressed the field equations of scalar-tensor gravity completely in terms of invariant quantities. In order to solve these field equations in a post-Newtonian approximation, we make use of the well known PPN formalism. Since we are dealing with different invariant metrics and their corresponding conformal frames, we briefly review the relevant parts of the PPN formalism for this situation. We start by introducing velocity orders in section~\ref{ssec:velorder}. These are used to define the PPN expansions of the scalar field in section~\ref{ssec:ppnscalar}, the invariant metrics in section~\ref{ssec:ppnmetric}, the energy-momentum tensor in section~\ref{ssec:ppnenmom} and the Ricci tensor in section~\ref{ssec:ppnricci}.

\subsection{Slow-moving source matter and velocity orders}
\label{ssec:velorder}

Starting point of the PPN formalism is the assumption of perfect fluid matter, for which the (Jordan frame) energy-stress tensor is given by
\ba
T^{\mathfrak{J}\,\m\n} = \l( \rho + \rho \Pi + p \r) u^\m u^\n + p g^{\mathfrak{J}\,\m\n} \,.
\ea
Since test particles fall on geodesics of the Jordan frame metric, we consider this as the `physical metric' and we define mass density $\rho$, pressure $p$ and specific internal energy $\Pi$ in this frame.
By $u^\m$ we denote the four-velocity, normalized such that $u^\mu u_\mu = -1$, where indices are raised and lowered using the Jordan frame metric $g^{\mathfrak{J}}_{\m\n}$.

We now consider the PPN framework to expand and solve the field equations up to the first post-Newtonian order. For this purpose we assume that the source matter is slow-moving, $v^i = u^i/u^0 \ll 1$. We use this assumption to expand all dynamical quantities in velocity orders $\mathcal{O}(n) \sim |\vec{v}|^n$.
Note that $\rho$ and $\Pi$ each contribute at order $\mathcal{O}(2)$, while $p$ contributes at $\mathcal{O}(4)$. The velocity terms $v^i$ are, obviously, of order $\mathcal{O}(1)$. We finally assume a quasi-static solution, where any time evolution is caused by the motion of the source matter. Hence, each time derivative $\partial_0 \sim \mathcal{O}(1)$ increases the velocity order of a term by one.

\subsection{PPN expansion of the scalar field}
\label{ssec:ppnscalar}

We now expand the scalar field around its cosmological background value $\Phi_0$ in terms of velocity orders,
\ba
\Phi = \Phi_0 + \phi
= \Phi_0 + \order{\phi}{2} + \order{\phi}{4} + \mathcal{O}{(6)}\,,
\ea
where $\order{\phi}{2}$ is of order $\mathcal{O}{(2)}$ and $\order{\phi}{4}$ is of order $\mathcal{O}{(4)}$. Other velocity orders either vanish due to conservation laws or are not relevant for the PPN calculation.
Any function of the scalar field $\mathcal{X}(\P)$ can then be expanded in a Taylor series as
\ba
\bs
\mathcal{X}(\P)
&= \mathcal{X}(\P_0) + \mathcal{X}'(\P_0) \phi + \f12 \mathcal{X}''(\P_0) \phi^2 + \mathcal{O}(6)
\\
&= \mathcal{X}(\P_0) + \mathcal{X}'(\P_0) \order{\phi}{2}
+ \l[ \mathcal{X}'(\P_0) \order{\phi}{4} + \f12 \mathcal{X}''(\P_0) \order{\phi}{2} \,\, \order{\phi}{2} \r]
+ \mathcal{O}{(6)} \,.
\es
\ea
For convenience, we denote the Taylor expansion coefficients, which are given by the values of the functions and their derivatives evaluated at the background value, in the form
$F \equiv \F(\Phi_0)$,
$F' \equiv \F'(\Phi_0)$,
$I_1 \equiv \I_1(\Phi_0)$,
$I_1' \equiv \I_1'(\Phi_0)$,
$I_1'' \equiv \I_1''(\Phi_0)$,
and similarly for all functions of the scalar field.

\subsection{PPN expansion of the metric tensors}
\label{ssec:ppnmetric}

In the next step, we assume that the Jordan frame metric, which governs the geodesic motion of test masses, is asymptotically flat, and can be expanded around a Minkowski vacuum solution in suitably chosen Cartesian coordinates. The expansion of the Jordan frame metric components up to the first post-Newtonian order is then given by
\begin{subequations}\label{eqn:metricjppn}
\begin{align}
g^{\mathfrak{J}}_{00} &= -1 + \order{h}{2}^{\mathfrak{J}}_{00} + \order{h}{4}^{\mathfrak{J}}_{00} + \mathcal{O}(6)\,,\\
g^{\mathfrak{J}}_{0i} &= \order{h}{3}^{\mathfrak{J}}_{0i} + \mathcal{O}(5)\,,\\
g^{\mathfrak{J}}_{ij} &= \delta_{ij} + \order{h}{2}^{\mathfrak{J}}_{ij} + \mathcal{O}(4)\,.
\end{align}
\end{subequations}
It can be shown that these are all relevant and non-vanishing components.
A similar expansion of the Einstein frame metric \(g^{\mathfrak{E}}_{\mu\nu}\) can be defined as
\begin{subequations}\label{eqn:metriceppn}
\begin{align}
I_1g^{\mathfrak{E}}_{00} &= -1 + \order{h}{2}^{\mathfrak{E}}_{00} + \order{h}{4}^{\mathfrak{E}}_{00} + \mathcal{O}(6)\,,\\
I_1g^{\mathfrak{E}}_{0i} &= \order{h}{3}^{\mathfrak{E}}_{0i} + \mathcal{O}(5)\,,\\
I_1g^{\mathfrak{E}}_{ij} &= \delta_{ij} + \order{h}{2}^{\mathfrak{E}}_{ij} + \mathcal{O}(4)\,.
\end{align}
\end{subequations}
The $I_1$'s on the left sides are required in order to satisfy \eqref{BETA equ: Einstein Jordan frame metric relation}.
The expansion coefficients in the two frames are then related by
\begin{subequations}
\begin{align}
\order{h}{2}^{\mathfrak{E}}_{00} &= \order{h}{2}^{\mathfrak{J}}_{00}
+ \frac{I_{1}'}{I_1}\order{\phi}{2}\,,\\
\order{h}{2}^{\mathfrak{E}}_{ij} &= \order{h}{2}^{\mathfrak{J}}_{ij}
- \frac{I_{1}'}{I_1}\order{\phi}{2} \delta_{ij}\,,\\
\order{h}{3}^{\mathfrak{E}}_{0i} &= \order{h}{3}^{\mathfrak{J}}_{0i}\,,\\
\order{h}{4}^{\mathfrak{E}}_{00} &= \order{h}{4}^{\mathfrak{J}}_{00}
+ \frac{I_{1}'}{I_1}\order{\phi}{4} + \frac{I_1I_{1}'' - 2I_{1}' I_{1}'}{2I_1^2}\order{\phi}{2} \, \order{\phi}{2}
- \frac{I_{1}'}{I_1}\order{\phi}{2} \,\order{h}{2}^{\mathfrak{J}}_{00}\,,
\end{align}
\end{subequations}
as one easily checks.
Conversely, one finds the inverse relations
\begin{subequations}
\label{BETA.equ:metric E to J frame}
\begin{align}
\order{h}{2}^{\mathfrak{J}}_{00} &= \order{h}{2}^{\mathfrak{E}}_{00}
- \frac{I_{1}'}{I_1}\order{\phi}{2} \,,\\
\order{h}{2}^{\mathfrak{J}}_{ij} &= \order{h}{2}^{\mathfrak{E}}_{ij}
+ \frac{I_{1}'}{I_1}\order{\phi}{2} \delta_{ij}\,,\\
\order{h}{3}^{\mathfrak{J}}_{0i} &= \order{h}{3}^{\mathfrak{E}}_{0i}\,,\\
\order{h}{4}^{\mathfrak{J}}_{00} &= \order{h}{4}^{\mathfrak{E}}_{00}
- \frac{I_{1}'}{I_1}\order{\phi}{4} - \frac{I_{1}''}{2I_1}\order{\phi}{2} \, \order{\phi}{2}
+ \frac{I_{1}'}{I_1}\order{\phi}{2} \, \order{h}{2}^{\mathfrak{E}}_{00}\,.
\end{align}
\end{subequations}

\subsection{PPN expansion of the energy-momentum tensors}
\label{ssec:ppnenmom}

We now come to the PPN expansion of the energy-momentum tensors. Here we restrict ourselves to displaying the expansion of the invariant energy-momentum tensor in the Einstein frame, since this is the frame we will be using for solving the field equations. It is related to the invariant Jordan frame energy-momentum tensor by
\(T^{\mathfrak{E}}_{\mu\nu} = \mathcal{I}_1T^{\mathfrak{J}}_{\mu\nu}\).
%
Its PPN expansion follows from the standard PPN expansion of the energy-momentum tensor in the Jordan frame~\cite{BETA.will.book} and is given by
\begin{subequations}
\begin{align}
T^{\mathfrak{E}}_{00} &= I_1\rho\left(1 + \frac{2 I_{1,A}}{I_1}\order{\phi}{2}^A - \order{h}{2}^{\mathfrak{E}}_{00} + v^2 + \Pi\right) + \mathcal{O}(6)\,,\\
T^{\mathfrak{E}}_{0i} &= - I_1\rho v_i + \mathcal{O}(5)\,,\\
T^{\mathfrak{E}}_{ij} &= I_1(\rho v_iv_j + p\delta_{ij}) + \mathcal{O}(6)\,.
\end{align}
\end{subequations}
Its trace, taken using the Einstein frame metric, has the PPN expansion
\begin{equation}
T^{\mathfrak{E}} = I_1^2\left(-\rho + 3p - \Pi\rho - 2 \frac{I_{1,A}}{I_1}\rho\order{\phi}{2}^A\right) \,.
\end{equation}
Consequently, the trace-reversed energy-momentum tensor is given by
\begin{subequations}
\begin{align}
\bar{T}^{\mathfrak{E}}_{00} &= I_1\rho\left(\frac{1}{2} + \frac{I_{1,A}}{I_1}\order{\phi}{2}^A - \frac{\order{h}{2}^{\mathfrak{E}}_{00}}{2} + v^2 + \frac{\Pi}{2} + \frac{3p}{2\rho}\right) + \mathcal{O}(6)\,,\\
\bar{T}^{\mathfrak{E}}_{0i} &= - I_1\rho v_i + \mathcal{O}(5)\,,\\
\bar{T}^{\mathfrak{E}}_{ij} &= I_1\rho\left[v_iv_j + \frac{\order{h}{2}^{\mathfrak{E}}_{ij}}{2} + \left(\frac{1}{2} + \frac{I_{1,A}}{I_1}\order{\phi}{2}^A + \frac{\Pi}{2} - \frac{p}{2\rho}\right)\delta_{ij}\right] + \mathcal{O}(6)\,.
\end{align}
\end{subequations}

\subsection{Invariant Ricci tensor}
\label{ssec:ppnricci}

Finally, we come to the PPN expansion of the Ricci tensor of the invariant Einstein metric. We will do this in a particular gauge, which is determined by the gauge conditions
\bsub
\ba
h^{\mathfrak{E}}_{ij,j} - h^{\mathfrak{E}}_{0i,0} - \frac{1}{2}h^{\mathfrak{E}}_{jj,i} + \frac{1}{2}h^{\mathfrak{E}}_{00,i} = 0 \,,
\\
h^{\mathfrak{E}}_{ii,0} = 2h^{\mathfrak{E}}_{0i,i} \,,
\ea
\esub
which will simplify the calculation. In this gauge, the components of the Ricci tensor to the orders that will be required are given by
\bsub
\ba
\order{R}{2}^{\mathfrak{E}}_{00} &= -\frac{1}{2}\triangle\order{h}{2}^{\mathfrak{E}}_{00}\,,
\\
\order{R}{2}^{\mathfrak{E}}_{ij} &= -\frac{1}{2}\triangle\order{h}{2}^{\mathfrak{E}}_{ij}\,,
\\
\order{R}{3}^{\mathfrak{E}}_{0i} &= -\frac{1}{2}\left(\triangle\order{h}{3}^{\mathfrak{E}}_{0i} + \frac{1}{2}\order{h}{2}^{\mathfrak{E}}_{jj,0i} - \order{h}{2}^{\mathfrak{E}}_{ij,0j}\right)\,,
\\
\order{R}{4}^{\mathfrak{E}}_{00} &= -\frac{1}{2}\triangle\order{h}{4}^{\mathfrak{E}}_{00} + \order{h}{3}^{\mathfrak{E}}_{0i,0i} - \frac{1}{2}\order{h}{2}^{\mathfrak{E}}_{ii,00} + \frac{1}{2}\order{h}{2}^{\mathfrak{E}}_{00,i}\left(\order{h}{2}^{\mathfrak{E}}_{ij,j} - \frac{1}{2}\order{h}{2}^{\mathfrak{E}}_{jj,i} - \frac{1}{2}\order{h}{2}^{\mathfrak{E}}_{00,i}\right) + \frac{1}{2}\order{h}{2}^{\mathfrak{E}}_{ij}\order{h}{2}^{\mathfrak{E}}_{00,ij}\,.
\ea
\esub
We now have expanded all dynamical quantities which appear in the field equations into velocity orders. By inserting these expansions into the field equations, we can perform a similar expansion of the field equations, and decompose them into different velocity orders. This will be done in the next section.

\section{Expanded field equations}
\label{sec Expanded field equations}
We will now make use of the PPN expansions displayed in the previous section and insert them into the field equations. This will yield us a system of equations, which are expressed in terms of the metric and scalar field perturbations that we aim to solve for. We start with the zeroth order field equations in section~\ref{ssec:eqns0}, which are the equations for the Minkowski background, and will give us conditions on the invariant potential \(\mathcal{I}_2\). We then proceed with the second order metric equation in section~\ref{ssec:eqnsh2}, the second order scalar equation in section~\ref{ssec:eqnsp2}, the third order metric equation in section~\ref{ssec:eqnsh3}, the fourth order metric equation in section~\ref{ssec:eqnsh4} and finally the fourth order scalar equation in section~\ref{ssec:eqnsp4}.

\subsection{Zeroth order metric and scalar equations}
\label{ssec:eqns0}

At the zeroth velocity order, the metric equations \eqref{equ: full metric field equation E-frame} are given by
\begin{equation}\label{eqn:h0mn}
-\kappa^2\frac{I_2}{I_1}\eta_{\mu\nu} = 0\,,
\end{equation}
which is satisfied only for \(I_2 = 0\), and hence restricts the choice of the invariant potential \(\mathcal{I}_2\).
At the same velocity order, the scalar equation reads
\begin{equation}\label{eqn:phi0}
-\frac{\kappa^2}{2}I_{2}'  = 0 \,,
\end{equation}
and is solved only by \(I_{2}' = 0\), so that we obtain another restriction on the allowed potential $\mathcal{I}_2$. In the following, we will only consider theories in which these conditions on $\mathcal{I}_2$ are satisfied.

\subsection{Second order metric $h^{\mathfrak{E}}_{00}$ and $h^{\mathfrak{E}}_{ij}$}
\label{ssec:eqnsh2}

At the second velocity order we find the $00$-metric field equation
\begin{equation}
\order{R}{2}^{\mathfrak{E}}_{00}
- \kappa^2\frac{I_2}{I_1}\order{h}{2}^{\mathfrak{E}}_{00}
+ \kappa^2\frac{I_{2}'}{I_1}\order{\phi}{2}^A
= \frac{\kappa^2}{2}I_1\rho \,.
\end{equation}
Inserting the expansion of the Ricci tensor shown in section~\ref{ssec:ppnricci} and using \(I_2 = 0\) and \(I_{2}' = 0\) we solve for \(\order{h}{2}^{\mathfrak{E}}_{00}\) and find the Poisson equation
\begin{equation}
\label{eqn:h200}
\triangle \order{h}{2}^{\mathfrak{E}}_{00} = -\kappa^2I_1\rho = -8\pi G\rho\,,
\end{equation}
where we introduced the Newtonian gravitational constant
\ba
\label{equ: Newtonian gravitational constant}
G = \frac{\kappa^2I_1}{8\pi}\,.
\ea
The $ij$-equations at the same order are given by
\ba
\order{R}{2}^{\mathfrak{E}}_{ij}
- \kappa^2\frac{I_2}{I_1}\order{h}{2}^{\mathfrak{E}}_{ij}
- \kappa^2\frac{I_{2}'}{I_1}\order{\phi}{2}^A\delta_{ij}
= \frac{\kappa^2}{2}I_1\rho\delta_{ij} \,,
\ea
which similarly reduces to
\ba
\label{eqn:h2ij}
\triangle\order{h}{2}^{\mathfrak{E}}_{ij} = -\kappa^2I_1\rho\delta_{ij} = -8\pi G\rho\delta_{ij}\,.
\ea
Note that the diagonal components $i=j$ satisfy the same equation~\eqref{eqn:h200} as \(\order{h}{2}^{\mathfrak{E}}_{00}\).

\subsection{Second order scalar field $\phi^A$}
\label{ssec:eqnsp2}

The second order scalar field equation is given by
\ba
I_1 F \triangle\order{\phi}{2}
- \frac{\kappa^2}{2}I_{2}''\order{\phi}{2}
= \frac{\kappa^2}{4}I_1I_{1}'\rho\,.
\ea
It is convenient to introduce the scalar field mass $m$ by
\ba
\label{equ: scalar mass}
m^2 &\equiv \frac{\kappa^2}{2} \f{1}{I_1 F} I_{2}''
\ea
and
\ba
k &= \frac{\kappa^2}{4} \f{1}{F} I_{1}' \,.
\ea
We assume that $m^2 > 0$, since otherwise the scalar field would be a tachyon.
Then, the second order scalar field equation takes the form of a screened Poisson equation,
\ba
\label{eqn:phi2}
\triangle\order{\phi}{2} - m^2 \order{\phi}{2} = k \rho\,.
\ea
We will see that $m$ can be interpreted as the mass of the scalar field, while $k$ is a measure for the non-minimal coupling of the scalar field at the linear level.  We finally remark that \(m\) is an invariant, while \(k\) transforms as a tangent vector to the real line of scalar field values~\cite{BETA.JarvInvariants2015}.

\subsection{Third order metric $h^{\mathfrak{E}}_{0i}$}
\label{ssec:eqnsh3}

The third order metric equation reads
\begin{equation}
\order{R}{3}^{\mathfrak{E}}_{0i} - \kappa^2\frac{I_2}{I_1}\order{h}{3}^{\mathfrak{E}}_{0i} = -\kappa^2I_1\rho v_i \,.
\end{equation}
Thus we can solve for the third order metric perturbation and obtain another Poisson equation,
\begin{equation}\label{eqn:h30i}
\triangle\order{h}{3}^{\mathfrak{E}}_{0i} = \order{h}{2}^{\mathfrak{E}}_{ij,0j} - \frac{1}{2}\order{h}{2}^{\mathfrak{E}}_{jj,0i} + 2\kappa^2I_1\rho v_i\,.
\end{equation}
Note that the source terms on the right hand side of this equation are given by time derivatives of other metric components and moving source matter, and hence vanish for static solutions and non-moving sources.

\subsection{Fourth order metric $h^{\mathfrak{E}}_{00}$}
\label{ssec:eqnsh4}

The fourth order metric field equation reads
\begin{equation}
\bs
\order{R}{4}^{\mathfrak{E}}_{00}
- \kappa^2\frac{I_2}{I_1}\order{h}{4}^{\mathfrak{E}}_{00}
+ \kappa^2\frac{I_{2}'}{I_1}\order{\phi}{4}
- \kappa^2\frac{I_{2}'}{I_1}\order{\phi}{2} \; \order{h}{2}^{\mathfrak{E}}_{00}
+ \frac{\kappa^2}{2}\frac{I_{2}''}{I_1}\order{\phi}{2} \; \order{\phi}{2}
\\
= \frac{\kappa^2}{2}I_1\rho\left(2\frac{I_{1}'}{I_1}\order{\phi}{2}
- \order{h}{2}^{\mathfrak{E}}_{00}
+ 2v^2 + \Pi + 3\frac{p}{\rho}\right)\,.
\es
\end{equation}
Solving for the fourth order metric perturbation then yields
\begin{equation}\label{eqn:h400}
\begin{split}
\triangle\order{h}{4}^{\mathfrak{E}}_{00}
&= 2\order{h}{3}^{\mathfrak{E}}_{0i,0i}
- \order{h}{2}^{\mathfrak{E}}_{ii,00}
+ \order{h}{2}^{\mathfrak{E}}_{00,i}\left(\order{h}{2}^{\mathfrak{E}}_{ij,j}
- \frac{1}{2}\order{h}{2}^{\mathfrak{E}}_{jj,i}
- \frac{1}{2}\order{h}{2}^{\mathfrak{E}}_{00,i}\right)
+ \order{h}{2}^{\mathfrak{E}}_{ij}\order{h}{2}^{\mathfrak{E}}_{00,ij}\\
&\phantom{=}+ \kappa^2\left(\frac{I_{2}''}{I_1}\order{\phi}{2} \; \order{\phi}{2}
- 2I_{1}'\order{\phi}{2} \rho
+ I_1\order{h}{2}^{\mathfrak{E}}_{00}\rho
- 2 I_1 v^2 \rho - I_1 \Pi \rho - 3 I_1 p \right)\,.
\end{split}
\end{equation}
Also this equation has the form of a Poisson equation.

\subsection{Fourth order scalar field $\phi^A$}
\label{ssec:eqnsp4}

Finally, for the scalar field we have the fourth order equation
\begin{multline}
I_1 F \triangle\order{\phi}{4}
- I_1 F \order{\phi}{2}_{,00}
- \frac{\kappa^2}{2}I_{2}''\order{\phi}{4}
- I_1 F \order{\phi}{2}_{,ij}\order{h}{2}^{\mathfrak{E}}_{ij}
+ I_1 F' \triangle\order{\phi}{2} \; \order{\phi}{2}
\\
+ \frac{I_1}{2} F' \order{\phi}{2}_{,i}\order{\phi}{2}_{,i}
+ \frac{I_1}{2} F  \order{\phi}{2}_{,i}\left(2\order{h}{2}^{\mathfrak{E}}_{ij,j}
- \order{h}{2}^{\mathfrak{E}}_{jj,i}
+ \order{h}{2}^{\mathfrak{E}}_{00,i}\right)
- \frac{\kappa^2}{4}I_{2}'''\order{\phi}{2} \; \order{\phi}{2} \\
= -\frac{\kappa^2}{4}\left[3I_1I_{1,A}p
- I_1 I_1' \Pi\rho
- (I_{1}' I_{1}'
+ I_1 I_1'')\order{\phi}{2} \rho \right] \,.
\end{multline}
Solving for the fourth order scalar perturbation then yields
\begin{equation}\label{eqn:phi4}
\begin{split}
\triangle\order{\phi}{4}
- m^2 \order{\phi}{4}
&= \order{\phi}{2}_{,00}
+ \order{\phi}{2}_{,ij} \order{h}{2}^{\mathfrak{E}}_{ij}
- \frac{1}{2}\order{\phi}{2}_{,i}\left(2\order{h}{2}^{\mathfrak{E}}_{ij,j}
- \order{h}{2}^{\mathfrak{E}}_{jj,i}
+ \order{h}{2}^{\mathfrak{E}}_{00,i}\right)
- \f{F'}{F} \left[ \triangle\order{\phi}{2} \; \order{\phi}{2}
+ \frac{1}{2} \order{\phi}{2}_{,i} \order{\phi}{2}_{,i}\right]\\
&\phantom{=}+ \frac{\kappa^2}{4} \f1F \left[\frac{I_{2}'''}{I_1}
\order{\phi}{2} \; \order{\phi}{2}
- 3 I_{1}' p + I_{1}' \Pi \rho + \left(\frac{({I_1}')^2}{I_1}
+ {I_1}'' \right) \order{\phi}{2} \rho\right]\,.
\end{split}
\end{equation}
This is again a screened Poisson equation, which contains the same mass parameter \(m\) as the second order scalar field equation~\eqref{eqn:phi2}.

These are all necessary equations in order to determine the relevant perturbations of the invariant Einstein frame metric and the scalar field. We will solve them in the next section, under the assumption of a massive scalar field, \(m > 0\), and a static, homogeneous, spherically symmetric source mass.


\section{Massive field and spherical source}
\label{sec Massive field and spherical source}
In the previous section we derived the gravitational field equations up to the required post-Newtonian order. We will now solve these field equations for the special case of a homogeneous, non-rotating spherical mass distribution. This mass distribution, as well as the corresponding ansatz for the PPN metric perturbation and the PPN parameters, are defined in section~\ref{ssec:homosphere}. We then solve the field equations by increasing order. The second order equations for the invariant Einstein frame metric and the scalar field are solved in sections~\ref{ssec:solh2} and~\ref{ssec:solp2}, while the corresponding fourth order equations are solved in sections~\ref{ssec:solh4} and~\ref{ssec:solp4}. From these solutions we read off the effective gravitational constant as well as the PPN parameters \(\gamma\) and \(\beta\) in section~\ref{sec PPN parameters}. A few limiting cases of this result are discussed in section~\ref{ssec:limits}.

\subsection{Ansatz for homogeneous, spherical mass source}
\label{ssec:homosphere}

In the following we consider a static sphere of radius $R$ with homogeneous rest mass density, pressure and specific internal energy, surrounded by vacuum. Its density \(\rho\), pressure \(p\) and specific internal energy \(\Pi\) are then given by
\ba\label{eqn:homosource}
\rho(r) =
\begin{cases}
	\rho_0 & \text{if } r \leq R\\
	0, & \text{if } r > R\\
\end{cases} \,,
\quad
p(r) =
\begin{cases}
	p_0 & \text{if } r \leq R\\
	0, & \text{if } r > R\\
\end{cases} \,,
\quad
\Pi(r) =
\begin{cases}
	\Pi_0 & \text{if } r \leq R\\
	0, & \text{if } r > R\\
\end{cases} \,,
\ea
where \(r\) is the radial coordinate and we use isotropic spherical coordinates. We further assume that the mass source is non-rotating and at rest with respect to our chosen coordinate system, so that the velocity \(v^i\) vanishes.

For the metric perturbation corresponding to this matter distribution, which is likewise spherically symmetric, we now use the ansatz
\begin{subequations}
\label{BETA.equ:PPN metric ansatz}
\begin{align}
\label{BETA.equ:PPN metric ansatz h200}
\order{h}{2}^{\mathfrak{J}}_{00} &= 2 G_\text{eff} U
\,,\\
\label{BETA.equ:PPN metric ansatz h2ij}
\order{h}{2}^{\mathfrak{J}}_{ij} &= 2 \gamma G_\text{eff} U \delta_{ij}
\,,\\
\label{BETA.equ:PPN metric ansatz h30i}
\order{h}{3}^{\mathfrak{J}}_{0i} &= 0
\,,\\
\label{BETA.equ:PPN metric ansatz h400}
\order{h}{4}^{\mathfrak{J}}_{00} &= -2 \beta G_\text{eff}^2 U^2
+ 2 G_\text{eff}^2 (1+3 \gamma-2 \beta) \Phi_2
+ G_\text{eff}(2\Phi_3 +6 \gamma \Phi_4)
\,.
\end{align}
\end{subequations}
Here \(U, \Phi_2, \Phi_3, \Phi_4\) denote the standard PPN potentials, which satisfy the Poisson equations~\cite{BETA.will.book}
\bsub
\label{BETA.equ: Poisson equ potentials}
\ba
\label{BETA.equ: Poisson equ U}
\triangle U &= - 4 \pi \rho \,,
\\
\label{BETA.equ: Poisson equ Phi_2}
\triangle \Phi_2 &= - 4 \pi U \rho \,,
\\
\label{BETA.equ: Poisson equ Phi_3}
\triangle \Phi_3 &= - 4 \pi \rho \Pi \,,
\\
\label{BETA.equ: Poisson equ Phi_4}
\triangle \Phi_4 &= - 4 \pi p \,.
\ea
\esub
For the homogeneous, spherically symmetric mass source we consider they are given by
\bsub
\ba
U(r)
&= \begin{cases}
- \f{M}{2 R^3}(r^2 - 3 R^2)
& \text{if } r \leq R
\\
\frac{M}{r}
& \text{if } r > R
\\
\end{cases} \,,
\\
\Phi_2 &=
	\begin{cases}
	\f{3 M^2}{40 R^6}(r^2 - 5 R^2)^2 & \text{if } r \leq R
	\\
	\f{6 M^2}{5 R r} & \text{if } r > R
	\\
	\end{cases} \,,
\\
\Phi_3 &=
	\begin{cases}
	-\f{M \Pi_0}{2 R^3} (r^2 - 3 R^2) & \text{if } r \leq R
	\\
	\f{M \Pi_0}{r} & \text{if } r > R
	\\
	\end{cases} \,,
\\
\Phi_4 &=
	\begin{cases}
	-\f{2\pi p_0}{3} (r^2 - 3 R^2) & \text{if } r \leq R
	\\
	\f{4 \pi p_0 R^3}{3 r} & \text{if } r > R
	\\
	\end{cases} \,,
\ea
\esub
where \(M = \frac{4\pi}{3}\rho_0R^3\) is the total mass. The metric ansatz~\eqref{BETA.equ:PPN metric ansatz} further depends on the effective gravitational constant \(G_{\text{eff}}\) and the PPN parameters \(\gamma\) and \(\beta\). These quantities, which are sufficient to describe the post-Newtonian limit of a fully conservative theory, i.e., a theory without preferred location or preferred frame effects, are determined by the particular theory under consideration. Note that these parameters are, in general, not constant, if one considers a massive scalar field, as we will do in the following.

We finally remark that in the ansatz~\eqref{BETA.equ:PPN metric ansatz} we have used the perturbations of the invariant Jordan frame metric \(g^{\mathfrak{J}}_{\mu\nu}\) defined in~\eqref{BETA equ: Jordan frame metric}. This choice is related to the fact that the matter coupling, and hence the geodesic motion of test particles from which the PPN parameters are determined, is given by \(g^{\mathfrak{J}}_{\mu\nu}\).

\subsection{Second order metric}
\label{ssec:solh2}

We start by solving the metric field equations at the second velocity order. Its temporal component~\eqref{eqn:h200} takes the form
\ba
\triangle \order{h}{2}^{\mathfrak{E}}_{00}
= \begin{cases}
-\frac{3 \text{I}_1 \kappa^2 M}{4 \pi  R^3}
& \text{if } r \leq R
\\
0
& \text{if } r > R
\\
\end{cases}\,.
\ea
The solution is given by
\ba
\order{h}{2}^{\mathfrak{E}}_{00}
= 2GU
= \begin{cases}
- \f{I_1 \k^2 M}{8 \pi R^3}(r^2 - 3 R^2)
& \text{if } r \leq R
\\
\frac{\text{I}_1 \kappa ^2 M}{4 \pi  r}
& \text{if } r > R
\\
\end{cases} \,.
\ea
Since the spatial metric equations~\eqref{eqn:h2ij} at the same order are identical to the temporal equation, except for a Kronecker symbol, their solution immediately follows as
\ba
\order{h}{2}^{\mathfrak{E}}_{ij} = \order{h}{2}^{\mathfrak{E}}_{00} \delta_{ij} \,.
\ea

\subsection{Second order scalar}
\label{ssec:solp2}

We then continue with the scalar field equation~\eqref{eqn:phi2} at the second velocity order, which reads
\ba
\left(\triangle - m^2 \right) \order{\phi}{2}
= \begin{cases}
\frac{3 I_1' \kappa^2 M}{16 \pi F R^3}
& \text{if } r \leq R
\\
0
& \text{if } r > R
\\
\end{cases} \,.
\ea
The solution is then given by.
\ba
\order{\phi}{2}
= \begin{cases}
-\frac{3 I_1' \kappa^2 M}{16 \pi F m^2 R^3}+\frac{3 e^{-m R} I_1' \kappa^2 M \
(1+m R) }{16 \pi F m^3 R^3} \f{\sinh (m r)}{r}
& \text{if } r \leq R
\\
-\frac{3 \kappa^2 M I_1' \left( e^{-m R}(1+m R) + e^{m R} (-1+m R)\right)}{32 \pi F m^3 R^3} \f{e^{-m r}}{r}
& \text{if } r > R
\\
\end{cases} \,.
\ea
Note that outside the source, the field is proportional to $\f{e^{-m r}}{r}$, i.e., it has the form of a Yukawa potential. Therefore, the parameter $m$ can be interpreted as the mass of the scalar field.

\subsection{Fourth order metric}
\label{ssec:solh4}

Since the only third order equations are trivially solved by \(\order{h}{3}^{\mathfrak{E}}_{0i} = 0\) in the case of a static, non-moving matter source, we continue directly with the metric field equation at the fourth velocity order. As it is rather lengthy, we give here only its generic form, while all appearing coefficients are stated explicitly in the appendix. This generic form reads
\ba
\label{BETA equ hE004 equation}
\triangle \order{h}{4}^{\mathfrak{E}}_{00}
= \begin{cases}
A_{h400}^{I1}
+\frac{A_{h400}^{I2}}{r^2}
+A_{h400}^{I3} r^2
+\frac{A_{h400}^{I4} e^{-m r}}{r}
+\frac{A_{h400}^{I5} e^{-2 m r}}{r^2}
+\frac{A_{h400}^{I6} e^{m r}}{r}
+\frac{A_{h400}^{I7} e^{2 m r}}{r^2}
& \text{if } r \leq R
\\
\frac{A_{h400}^{E1}}{r^4}
+\frac{A_{h400}^{E2} e^{-2 m r}}{r^2}
& \text{if } r > R
\\
\end{cases}\,.
\ea
Also its solution is lengthy, and so we proceed in the same fashion to display only its generic form here, which is given by
\bsub
\ba
\bs
\label{BETA equ hE004 solution int}
\order{h}{4}^{\mathfrak{E}}_{00} (r \leq R)
&= B_{h400}^{I1}
+B_{h400}^{I2} r^2
+B_{h400}^{I3} r^4
+\frac{B_{h400}^{I4} e^{-m r}}{r}
+\frac{B_{h400}^{I5} e^{-2 m r}}{r}
+\frac{B_{h400}^{I6} e^{m r}}{r}
\\
&\phantom{=}+\frac{B_{h400}^{I7} e^{2 m r}}{r}
+B_{h400}^{I8} \mathrm{Ei}(-2 m r)
+B_{h400}^{I9} \mathrm{Ei}(2 m r)
+B_{h400}^{I10} \ln\l(\f{r}{R}\r) \,,
\es
\\
\label{BETA equ hE004 solution ext}
\order{h}{4}^{\mathfrak{E}}_{00} (r > R)
&=\frac{B_{h400}^{E1}}{r}
+\frac{B_{h400}^{E2}}{r^2}
+\frac{B_{h400}^{E3} e^{-2 m r}}{r}
+B_{h400}^{E4} \mathrm{Ei}(-2 m r) \,,
\ea
\esub
where $\mathrm{Ei}$ is the exponential integral defined as
\ba
\mathrm{Ei}(x) = -\fint_{-x}^{\infty}\frac{e^{-t}}{t}dt\,,
\ea
with $\fint$ denoting the Cauchy principal value of the integral. The values of the coefficients can be found in the appendix \ref{app coefficients fourth order metric}.
Note that $B_{h400}^{I8}=B_{h400}^{I9}$ and thus the exponential integral terms can be written more compactly as
\ba
B_{h400}^{I8} \mathrm{Ei}(-2 m r)+B_{h400}^{I9} \mathrm{Ei}(2 m r)
= 2 B_{h400}^{I8} \mathrm{Chi}(2 m r) \,,
\ea
where we used $\mathrm{Chi}$ for the hyperbolic cosine integral
\ba
\mathrm{Chi}(x) = \frac{\mathrm{Ei}(x) + \mathrm{Ei}(-x)}{2} = \upgamma + \ln x + \int_0^x\frac{\cosh t - 1}{t}dt\,,
\ea
and $\upgamma$ is Euler's constant.

\subsection{Fourth order scalar}
\label{ssec:solp4}

The final equation we must solve is the scalar field equation at the fourth velocity order, since the fourth order scalar field \(\order{\phi}{4}\) enters the Jordan frame metric perturbation \(\order{h}{4}^{\mathfrak{J}}_{00}\), from which the PPN parameter \(\beta\) is read off. This equation is similarly lengthy, and so also here we restrict ourselves to displaying only the generic form, which reads
\bsub
\ba
\bs
\label{BETA equ phi4 equation int}
\l( \triangle - m^2 \r) \order{\phi}{4}(r \leq R) &=
A_{\phi 4}^{I1}
+\frac{A_{\phi 4}^{I2}}{r^2}
+\frac{A_{\phi 4}^{I3}}{r^4}
+\frac{A_{\phi 4}^{I4} e^{-m r}}{r}
+\frac{A_{\phi 4}^{I5} e^{-2 m r}}{r^2}
+\frac{A_{\phi 4}^{I6} e^{-2 m r}}{r^3}\\
&\phantom{=}+\frac{A_{\phi 4}^{I7} e^{-2 m r}}{r^4}
+\frac{A_{\phi 4}^{I8} e^{m r}}{r}
+\frac{A_{\phi 4}^{I9} e^{2 m r}}{r^2}
+\frac{A_{\phi 4}^{I10} e^{2 m r}}{r^3}
+\frac{A_{\phi 4}^{I11} e^{2 m r}}{r^4}\\
&\phantom{=}+A_{\phi 4}^{I12} e^{-m r} r
+A_{\phi 4}^{I13} e^{m r} r \,,
\es
\\
\bs
\label{BETA equ phi4 equation ext}
\l( \triangle - m^2 \r) \order{\phi}{4}(r > R) &=
\frac{A_{\phi 4}^{E1} e^{-m r}}{r^2}
+\frac{A_{\phi 4}^{E2} e^{-2 m r}}{r^2}
+\frac{A_{\phi 4}^{E3} e^{-2 m r}}{r^3}
+\frac{A_{\phi 4}^{E4} e^{-2 m r}}{r^4}\,.
\es
\ea
\esub
The generic form of the solution then follows as
\bsub
\ba
\bs
\label{BETA equ phi4 solution int}
\phi_4(r \leq R)
&= B_{\phi 4}^{I1}
+\frac{B_{\phi 4}^{I2}}{r^2}
+\frac{B_{\phi 4}^{I3} e^{-m r}}{r}
+\frac{B_{\phi 4}^{I4} e^{-2 m r}}{r^2}
+\frac{B_{\phi 4}^{I5} e^{m r}}{r}
+\frac{B_{\phi 4}^{I6} e^{2 m r}}{r^2}
+B_{\phi 4}^{I7} e^{-m r}\\
&\phantom{=}+B_{\phi 4}^{I8} e^{-m r} r
+B_{\phi 4}^{I9} e^{-m r} r^2
+B_{\phi 4}^{I10} e^{m r}
+B_{\phi 4}^{I11} e^{m r} r
+B_{\phi 4}^{I12} e^{m r} r^2\\
&\phantom{=}+\frac{B_{\phi 4}^{I13} e^{-m r} \mathrm{Ei}(-m r)}{r}
+\frac{B_{\phi 4}^{I14} e^{m r} \mathrm{Ei}(-m r)}{r}
+\frac{B_{\phi 4}^{I15} e^{m r} \mathrm{Ei}(-3 m r)}{r}\\
&\phantom{=}+\frac{B_{\phi 4}^{I16} e^{m r} \mathrm{Ei}(m r)}{r}
+\frac{B_{\phi 4}^{I17} e^{-m r} \mathrm{Ei}(m r)}{r}
+\frac{B_{\phi 4}^{I18} e^{-m r} \mathrm{Ei}(3 m r)}{r} \,,
\es
\\
\bs
\label{BETA equ phi4 solution ext}
\phi_4(r > R) &=
\frac{B_{\phi 4}^{E1} e^{-m r}}{r}
+\frac{B_{\phi 4}^{E2} e^{-2 m r}}{r^2}
+\frac{B_{\phi 4}^{E3} e^{-m r} \mathrm{Ei}(-m r)}{r}
+\frac{B_{\phi 4}^{E4} e^{m r} \mathrm{Ei}(-2 m r)}{r}\\
&\phantom{=}+\frac{B_{\phi 4}^{E5} e^{m r} \mathrm{Ei}(-3 m r)}{r}
+\frac{B_{\phi 4}^{E6} e^{-m r} \ln\l(\f{r}{R}\r)}{r} \,.
\es
\ea
\esub
The coefficients can be found in the appendix \ref{app coefficients fourth order scalar}.

\subsection{PPN parameters}
\label{sec PPN parameters}

We now have solved the field equations which determine all terms that enter the Jordan frame metric, and hence contribute to the PPN parameters \(\gamma\) and \(\beta\). The Jordan frame metric is then obtained by inserting the solutions obtained before into the relation~\eqref{BETA.equ:metric E to J frame} between the different invariant metrics.
Using the metric ansatz~\eqref{BETA.equ:PPN metric ansatz h200} we find the effective gravitational `constant'
\ba
G_\text{eff}(r)
= \f{\order{h}{2}^{\mathfrak{J}}_{00}}{2 U}
= \begin{cases}
G \l[ 1 + 3
\f{\sinh(mr)(1+m R)e^{-mR} - 2 m r }{(2\omega + 3) m^3 r (r^2 - 3 R^2)} \r]
& \text{if } r \leq R
\\
G \l[ 1 + 3
\f{ mR\cosh(mR) - \sinh(mR) }{(2\omega + 3) m^3 R^3}e^{-mr} \r]
& \text{if } r > R
\\
\end{cases}\,.
\ea
Here we have introduced the abbreviation
\ba
\omega = 2F\frac{I_1^2}{I_1'^2} - \frac{3}{2}\,,
\ea
which is invariant under reparametrizations of the scalar field and chosen such that it agrees with the parameter $\omega$ in case of the Jordan-Brans-Dicke theory~\cite{BETA.Jordan:1959eg,BETA.Brans:1961sx}. In the next step, the PPN parameter $\gamma$ is obtained from the metric ansatz~\eqref{BETA.equ:PPN metric ansatz h2ij} giving
\ba
\gamma(r) = \f{\order{h}{2}^{\mathfrak{J}}_{ii}}{2 G_\text{eff} U} =
\begin{cases}
1+\frac{12 \left[2 e^{m (r+R)} m r-\left(-1+e^{2 m r}\right) (1+m R)\right]}{6 \left(-1+e^{2 m r}\right) (1+m R)+2 e^{m (r+R)} m r \left[-6 + (2\omega + 3) m^2 \left(r^2-3 R^2\right)\right]}
& \text{if } r \leq R
\\
1-\l(\frac{1}{2}+\frac{(2\omega + 3) m^3 R^3 e^{m r}}{6 [m R \cosh (m R)-\sinh (m R)]}\r)^{-1}
& \text{if } r > R
\\
\end{cases} \,.
\ea
Finally, the PPN parameter $\beta$ is obtained from the ansatz~\eqref{BETA.equ:PPN metric ansatz h400}, and hence can be obtained from
\ba
\beta(r)
= -\frac{\order{h}{4}^{\mathfrak{J}}_{00}
- 2 G_\text{eff}[ (1+3\gamma)G_\text{eff}\Phi_2 + \Phi_3 + 3 \gamma \Phi_4 ]}{2 G_\text{eff}^2 (U^2 + 2 \Phi_2)} \,.
\ea
Due to the even more lengthy, and practically irrelevant solution for \(\beta\) inside the source, we omit this part of the solution. The solution outside the source, $r>R$, takes the generic form
\ba
\bs
\label{BETA.equ: beta}
\beta(r>R)
&= \l[ \l(\f{1}{r^2} + \f1r \f{12}{5R} \r) \l(1 + e^{-m r} \f{C_{\b}^{E4}}{2} \r)^2 \r]^{-1}
\Bigg[\f{C_{\b}^{E1}}{r}
+ \f{C_{\b}^{E2}}{r^2}
+ C_{\b}^{E3} \f{e^{-m r}}{r}
+ C_{\b}^{E4} \f{e^{-m r}}{r^2} \\
&\qquad\qquad+ C_{\b}^{E5} \f{e^{-2 m r}}{r}
+ C_{\b}^{E6} \f{e^{-2 m r}}{r^2}
+ C_{\b}^{E7} \f{e^{-  m r}}{r} \mathrm{Ei}{(-m r)}
+ C_{\b}^{E8} \f{e^{m r}}{r} \mathrm{Ei}{(-2 m r)}\\
&\qquad\qquad+ C_{\b}^{E9} \f{e^{m r}}{r} \mathrm{Ei}{(-3 m r)}
+ C_{\b}^{E10} \mathrm{Ei}{(-2 m r)}
+ C_{\b}^{E11} \f{e^{-m r}}{r} \ln\l(\f{r}{R}\r)
\Bigg] \,.
\es
\ea
The values of the coefficients can be found in the appendix \ref{app PPN Beta}, where we further introduce the abbreviations
\ba
\sigma = \frac{2F(I_1I_1'' + I_1'^2) - F'I_1I_1'}{2F^2I_1^2}\,, \quad \mu = \kappa^2I_1'\frac{2FI_2''' - 3F'I_2''}{4F^3I_1^2}\,.
\ea
Both $\gamma$ and $\beta$ depend only on the parameters $m, \omega, \mu, \sigma$ of the theory, which are invariant both under conformal transformations and redefinitions of the scalar field, and on the radius of the sphere $R$. As expected, it is independent of $M,\Pi_0,p_0$, which are absorbed in the metric potentials, and characterize the source only.

\subsection{Limiting cases}
\label{ssec:limits}

We finally discuss a number of physically relevant limiting cases. We start this discussion with the massless limit, i.e., the case of a vanishing potential $\U \ra 0$, corresponding to $\I_2 \ra 0$. This limit is achieved by successively applying to our result the limits $\mu \ra 0$ and $m \ra 0$. For $\gamma$, which does not depend on $\mu$, we obtain the limit
\ba
\gamma(m \ra 0) = \frac{\omega + 1}{\omega + 2} = \frac{4FI_1^2 - I_1'^2}{4FI_1^2 + I_1'^2}\,.
\ea
For $\beta$ we find the limit
\ba
\bs
\beta(\mu \ra 0, m \ra 0) &= \frac{(2\omega + 3)\sigma - 8}{16(\omega + 2)^2}\\
&= 1 - \l( 1 + \f{1}{4 F} \f{I_1'^2}{I_1^2} \r)^{-2}
\l( \f{F'}{32 F^3} \f{I_1'^3}{I_1^3} + \f{1}{16 F^2} \f{I_1'^4}{I_1^4}
- \f{1}{16 F^2} \f{I_1'^2 I_1''}{I_1^3} \r) \,.
\es
\ea
These limits correspond to the result found in \cite{BETA.KuuskInvariantsMSTG2016}, if reduced to the case of a single scalar field.

Another interesting case is given by the large interaction distance limit. In the limit $r \ra \infty$ we obtain
\ba
\gamma(r \ra \infty) = 1 \,,
\ea
and
\ba
\bs
\beta(r \ra \infty)
&= \f{5 C_{\b}^{E1} R}{12 I_1^2} \\
&= 1 + 5\frac{\left[39+m^2 R^2 (20 m R - 33)\right] -3 (1+m R) [13+m R (13+2 m R)]e^{-2 m R} }{16 (2\omega + 3) m^5 R^5} \,.\label{eqn:betainf}
\es
\ea
Note that it does not take the GR value $1$ as one might expect. This is due to the fact that the finite self-energy of the extended mass source influences $\beta$. If in addition we take the limit $R\ra \infty$, we find that indeed $\beta$ goes to the GR value $1$,
\ba
\beta(r \ra \infty, R \ra \infty) = 1 \,.
\ea
Note that first we have to take the limit $r \ra \infty$, since the solution we used here is valid only in the exterior region $r > R$, and the limit \(R \to \infty\) would otherwise be invalid. We finally remark that the same limit \(\beta \to 1\) is also obtained for \(m \to \infty\), which becomes clear from the fact that \(m\) always appears multiplied by either \(r\) or \(R\).

This concludes our section on the static spherically symmetric solution we discussed here. We can now use our results for \(\beta\) and \(\gamma\) and compare them to Solar System measurements of these PPN parameters. We will do so in the next section.

\section{Comparison to observations}
\label{sec Comparison to observations}
In the preceding sections we have derived expressions for the PPN parameters \(\beta\) and \(\gamma\). We have seen that they depend on the radius \(R\) of the gravitating source mass, the interaction distance \(r\) and constant parameters \(m, \omega, \mu, \sigma\), which characterize the particular scalar-tensor theory under consideration and are invariant under both conformal transformations of the metric and redefinitions of the scalar field. We now compare our results to observations of the PPN parameters in the Solar System, in order to obtain bounds on the theory parameters.

\begin{figure}[hbtp]
\centering
\includegraphics[width=100mm]{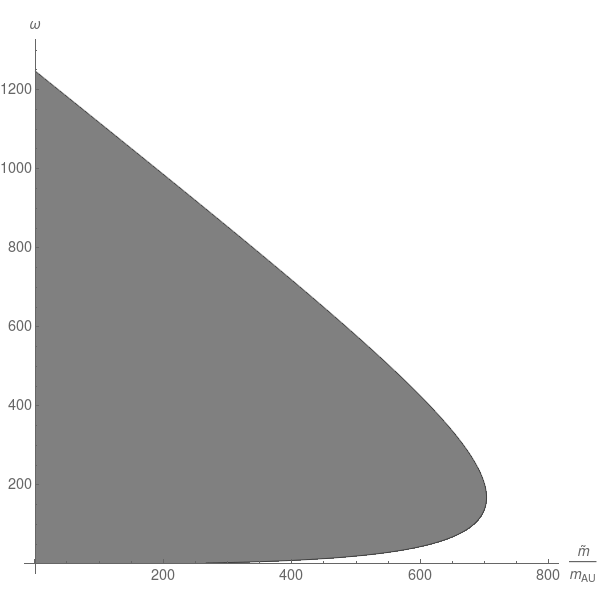}
\caption{Region excluded by VLBI measurements.}
\label{fig:vlbi}
\end{figure}

\begin{figure}[hbtp]
\centering
\includegraphics[width=100mm]{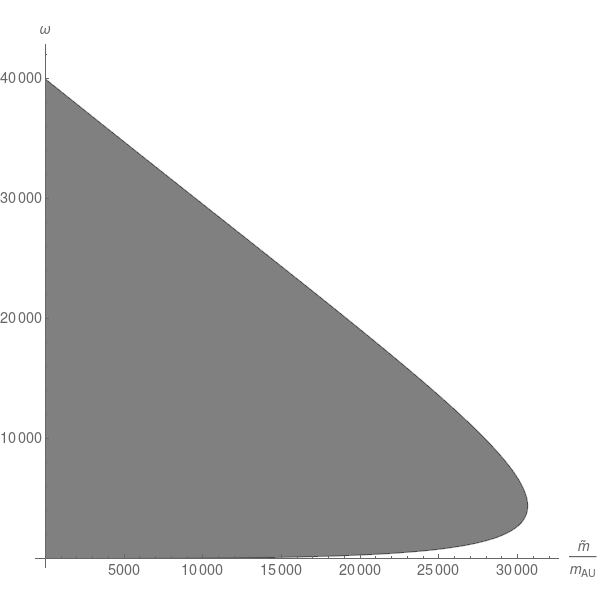}
\caption{Region excluded by Cassini tracking.}
\label{fig:cassini}
\end{figure}

\begin{figure}[hbtp]
\centering
\includegraphics[width=100mm]{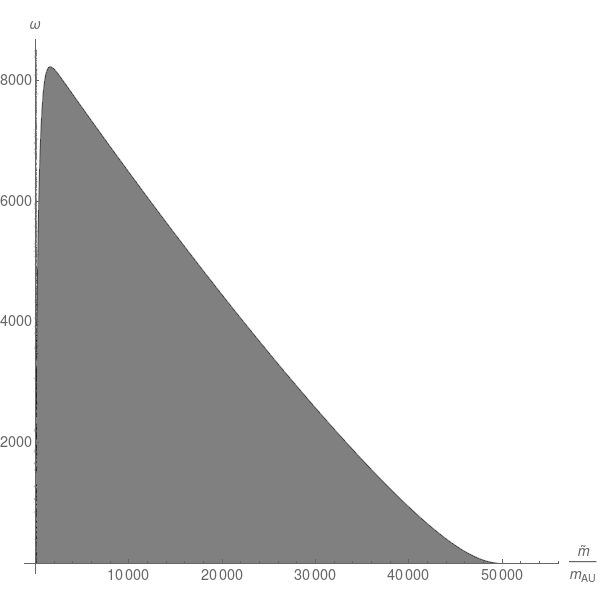}
\caption{Region excluded by the perihelion shift of Mercury.}
\label{fig:mercury}
\end{figure}

\begin{figure}[hbtp]
\centering
\includegraphics[width=100mm]{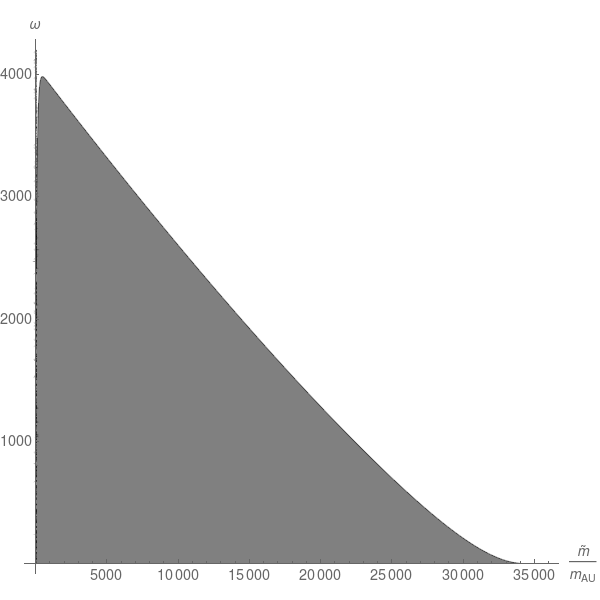}
\caption{Region excluded by lunar laser ranging.}
\label{fig:llr}
\end{figure}

\begin{figure}[hbtp]
\centering
\includegraphics[width=100mm]{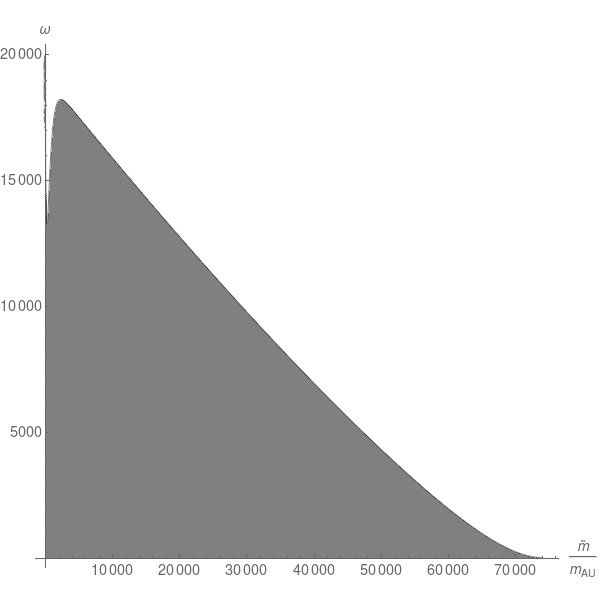}
\caption{Region excluded by the ephemeris data set INPOP13a.}
\label{fig:inpop}
\end{figure}

In the following we will not consider the parameters \(\mu\) and \(\sigma\), and set them to \(0\) in our calculations, as they correspond to higher derivatives of the invariant functions \(\mathcal{I}_1\) and \(\mathcal{I}_2\). Restricting our discussion to the parameters \(m\) and \(\omega\) will further allow us to plot exclusion regions which we can compare to previous results~\cite{BETA.HohmannPPN2013,BETA.HohmannPPN2013E,BETA.SchaererPPN2014}. To be compatible with the plots shown in these articles, we display the rescaled mass \(\tilde{m} = m/\sqrt{2\omega + 3}\) measured in inverse astronomical units \(m_{\mathrm{AU}} = 1\mathrm{AU}^{-1}\) on the horizontal axis. Regions that are excluded at a confidence level of \(2\sigma\) are shown in gray. In particular, we consider the following experiments:

\begin{itemize}
\item
The deflection of pulsar signals by the Sun has been measured using very long baseline interferometry (VLBI)~\cite{BETA.Fomalont:2009zg}. From this \(\gamma\) has been determined to satisfy \(\gamma - 1 = (-2 \pm 3) \cdot 10^{-4}\). The radar signals were passing by the Sun at an elongation angle of 3\textdegree, and so we will assume a gravitational interaction distance of \(r \approx 5.23 \cdot 10^{-2}\mathrm{AU}\). The region excluded from this measurement is shown in Fig.~\ref{fig:vlbi}.

\item
The most precise value for \(\gamma\) has been obtained from the time delay of radar signals sent between Earth and the Cassini spacecraft on its way to Saturn~\cite{BETA.Bertotti:2003rm}. The experiment yielded the value \(\gamma - 1 = (2.1 \pm 2.3) \cdot 10^{-5}\). The radio signals were passing by the Sun at a distance of \(1.6\) solar radii or \(r \approx 7.44 \cdot 10^{-3}\mathrm{AU}\). The excluded region, shown in Fig.~\ref{fig:cassini}, agrees with our previous findings~\cite{BETA.HohmannPPN2013,BETA.HohmannPPN2013E}.

\item
The classical test of the parameter \(\beta\) is the perihelion precession of Mercury~\cite{BETA.Will:2014kxa}. Its precision is limited by other contributions to the perihelion precession, most importantly the solar quadrupole moment \(J_2\). The current bound is \(\beta - 1 = (-4.1 \pm 7.8) \cdot 10^{-5}\). As the gravitational interaction distance we take the semi-major axis of Mercury, which is \(r \approx 0.387\mathrm{AU}\). We obtain the excluded region shown in Fig.~\ref{fig:mercury}. Note that for small values of \(\omega\) we obtain a tighter bound on the scalar field mass than from the Cassini tracking experiment, despite the larger interaction distance \(r\). This can be explained by the fact that the main contribution to \(\beta\) comes from a modification of the gravitational self-energy of the source mass, which is independent of the interaction distance, and depends only on the radius of the gravitating body.

\item
A combined bound on \(\beta\) and \(\gamma\) has been obtained from lunar laser ranging experiments searching for the Nordtvedt effect, which would cause a different acceleration of the Earth and the Moon in the solar gravitational field~\cite{BETA.Hofmann:2010}. In fully conservative theories with no preferred frame effects, such as scalar-tensor gravity, the Nordtvedt effect depends only on the PPN parameters \(\beta\) and \(\gamma\). The current bound is \(4\beta - \gamma - 3 = (0.6 \pm 5.2) \cdot 10^{-4}\). Since the effect is measured using the solar gravitational field, the interaction distance is \(r = 1\mathrm{AU}\). The excluded region is shown in Fig.~\ref{fig:llr}.

\item
A more recent measurement of both \(\beta\) and \(\gamma\) with higher precision has been obtained using combined ephemeris data and the Mercury flybys of the Messenger spacecraft in the INPOP13a data set~\cite{BETA.Verma:2013ata}. From these observations, combined bounds in the two-dimensional parameter space spanned by \(\beta\) and \(\gamma\) can be obtained, as well as bounds on the individual parameters by fixing one of them to its GR value. Since we have determined both parameters in our calculation, we do not perform such a fixing here, and use the full parameter space instead. From the 25\% residuals one finds a bounding region that can be approximated as
\begin{equation}
\left[(\beta - 1) - 0.2 \cdot 10^{-5}\right]^2 + \left[(\gamma - 1) + 0.3 \cdot 10^{-5}\right]^2 \leq \left(2.5 \cdot 10^{-5}\right)^2\,.
\end{equation}
Note that in this case one cannot easily define an interaction distance \(r\), since ephemeris from objects across the Solar System has been used. However, we may use the fact that for \(mr \gg 1\) the PPN parameters approach their limiting values \(\gamma \to 1\) and~\eqref{eqn:betainf}, so that the dominant effect is determined by the modified gravitational self-energy of the Sun. The excluded region under this assumption is shown in Fig.~\ref{fig:inpop}. One can see that for small values of \(\omega\) one obtains a bound on the scalar field mass which is approximately twice as large as the bound obtained from Cassini tracking and lunar laser ranging.
\end{itemize}

Our results must be taken with care, since they are based on a number of assumptions and simplifications. Most importantly, we have calculated the PPN parameters under the assumption of a homogeneous, non-rotating, spherical gravitational source. This is only a very crude approximation for the Sun, whose density decreases with growing distance from its center. A full treatment of the post-Newtonian limit of a non-homogeneous body would be required to improve on this assumption. However, since a larger amount of matter is located closer to the center of the Sun, hence increasing its gravitational self-energy and decreasing the effective radius \(R\), one might expect that the effect on \(\beta\) will be even larger in such a full treatment.

As another simplification we have assumed that experiments based on electromagnetic waves passing by the Sun can be described by a single effective interaction distance. A rigorous treatment would involve an explicit calculation of the wave trajectory~\cite{BETA.Devi:2011zz,BETA.Deng:2016moh}. However, this affects only the VLBI and Cassini measurements of \(\gamma\), while the measurements of \(\beta\), which are less dependent on the interaction distance, are unaffected.


\section{Conclusion}
\label{sec Conclusion}
We have calculated the PPN parameters \(\gamma\) and \(\beta\) of scalar-tensor gravity with a general potential for a homogeneous, spherical mass source. For our calculation we have used a formalism which is manifestly invariant under both conformal transformations of the metric and redefinitions of the scalar field. The result we have obtained depends on four constant parameters of the theory under consideration, which are derived from the invariant functions that characterize the theory. Further, the result also depends on the radius \(R\) of the gravitating mass source and the interaction distance \(r\) at which the PPN parameters are measured. We have finally compared our results to a number of measurements in the Solar System and derived bounds on two of the four constant theory parameters.

Our results improve on previous work in which we assumed a point-like mass source~\cite{BETA.HohmannPPN2013,BETA.HohmannPPN2013E,BETA.SchaererPPN2014}. We have seen that \(\gamma\) receives a correction which depends on the source mass radius, but retains the large distance limit \(\gamma \to 1\) for \(r \to \infty\). In contrast, \(\beta\) receives a modification also in the large distance limit. This is explained by a modified gravitational self-energy of the source mass, which influences its gravitational effects also at large distances, and which has been neglected for the point mass. As a result, measurements of \(\beta\) at an interaction distance which is large compared to the radius of the source mass, \(r \gg R\), are significantly more sensitive to modifications of GR by a massive scalar field than measurements of \(\gamma\) at the same interaction distance. We have shown this in particular for measurements of \(\beta\) using lunar laser ranging and planetary ephemeris, where the interaction distance is in the order of astronomical units, and which yield bounds on the scalar field mass comparable to or even better that the bound obtained from the Cassini tracking experiment, with an interaction distance in the order of the solar radius. Our work suggests that measurements of \(\beta\) in the gravitational field of smaller, more compact objects could yield even stricter bounds.

Of course also our assumption of a spherically symmetric and homogeneous source mass is still only an approximation. Further improvement of our results would require a weakening of this assumption, and considering the density profile of the gravitating source mass. Such a calculation would have to be done numerically. While we have provided all necessary equations in this article, we leave performing such a calculation for future work.

Finally, it is also possible to extend our results to more general or related theories. A straightforward generalization is given by considering multiple scalar fields, for which \(\gamma\) has been calculated for a point mass source~\cite{BETA.HohmannGamma2016}, or by allowing derivative coupling as in Horndeski gravity, where it would lead to a similar improvement on previous calculations of \(\gamma\) and \(\beta\)~\cite{BETA.Hohmann:2015kra}. Another possibility is to consider massive bimetric gravity, where GR is augmented by a massive tensor degree of freedom instead of a scalar field, and a similar result on \(\gamma\) for a point mass can be improved and extended to \(\beta\)~\cite{BETA.Hohmann:2017uxe}.


\section*{Acknowledgments}
The authors thank Sofya Labazova for pointing out an error in a previous calculation.
MH gratefully acknowledges the full financial support of the Estonian Research Council through the Startup Research Grant PUT790 and the European Regional Development Fund through the Center of Excellence TK133 ``The Dark Side of the Universe''.
MH and AS acknowledge support from the Swiss National Science Foundation.
This article is based upon work from COST Action CANTATA, supported by COST (European Cooperation in Science and Technology).


\pagebreak

\appendix

\section{Coefficients for the field equations}
\label{app coefficients}

In this appendix we list the constant expansion coefficients which appear in the fourth order PPN equations and their solutions.

\subsection{Fourth order metric}
\label{app coefficients fourth order metric}

The coefficients for the $00$-metric equation \eqref{BETA equ hE004 equation} are
\bsub
\ba
A_{h400}^{I1} &=
\frac{45 I_1'^2 \kappa^4 M^2 }{128 \pi^2 F m^2 R^6}
-3 I_1 \kappa^2 \text{p}_0
-\frac{3 I_1 \kappa^2 M  \Pi _0}{4 \pi  R^3}
\\
A_{h400}^{I2} &= -\frac{9 e^{-2 m R} I_1'^2 \kappa^4 M^2 (1+m R)^2 }{256 \pi^2 F m^4 R^6}
\\
A_{h400}^{I3} &= -\frac{I_1^2 \kappa^4 M^2}{16 \pi ^2 R^6}
\\
A_{h400}^{I4} &= -\frac{6 e^{m R} m}{1+m R} A_{h400}^{I2}
\\
A_{h400}^{I5} &= -\f12 A_{h400}^{I2}
\\
A_{h400}^{I6} &= - A_{h400}^{I4}
\\
A_{h400}^{I7} &= A_{h400}^{I5}
\\
A_{h400}^{E1} &= R^6 A_{h400}^{I3}
\\
A_{h400}^{E2} &= \frac{9 I_1'^2 \kappa^4 M^2 \left(e^{-m R} (1+m R)+e^{m R} (-1+m R)\right)^2 }{512 \pi^2 F m^4 R^6} \,.
\ea
\esub
\newpage
For the interior part of the solution \eqref{BETA equ hE004 solution int} they are
\bsub
\ba
\bs
B_{h400}^{I1} &= \frac{3 \kappa^4 M^2 I_1^2}{64 \pi ^2 R^2}
+\frac{3}{2} I_1 R^2 \kappa^2 \text{p}_0
+\frac{3 I_1 \kappa^2 M \Pi _0}{8 \pi R}
\\
&\phantom{=}+\frac{9 \kappa^4 M^2 I_1'^2}{256 \pi^2 F m^4 R^6}
\Bigg(
e^{-2 m R}(7+8 m R+m^2 R^2)
+ \left(6+6 m R-5 m^2 R^2\right) \\
&\phantom{=}+\frac{1}{2} (-1+m R) \left(2+2 m R+e^{2 m R} (-1+m R)\right) \mathrm{Ei}(-2 m R) \\
&\phantom{=}- \frac{1}{2} (1+m R)^2 e^{-2 m R} \mathrm{Ei}(2 m R)
\Bigg)
\es
\\
B_{h400}^{I2} &= -\frac{1}{2} I_1 \kappa^2 \text{p}_0
+\frac{15 I_1'^2 \kappa^4 M^2}{256 \pi^2 F m^2 R^6}-\frac{I_1 \kappa^2 M \Pi _0}{8 \pi R^3}
\\
B_{h400}^{I3} &= -\frac{I_1^2 \kappa^4 M^2}{320 \pi ^2 R^6}
\\
B_{h400}^{I4} &= \frac{27 e^{-m R} I_1'^2 \kappa^4 M^2 (1+m R) }{128 \pi^2 F m^5 R^6}
\\
B_{h400}^{I5} &= \frac{9 e^{-2 m R} I_1'^2 \kappa^4 M^2 (1+m R)^2}{1024 \pi^2 F m^5 R^6}
\\
B_{h400}^{I6} &= - B_{h400}^{I4}
\\
B_{h400}^{I7} &= - B_{h400}^{I5}
\\
B_{h400}^{I8} &= 2 m B_{h400}^{I5}
\\
B_{h400}^{I9} &= 2 m B_{h400}^{I5}
\\
B_{h400}^{I10} &= -4 m B_{h400}^{I5}
\ea
\esub
and for the exterior part \eqref{BETA equ hE004 solution ext}
\bsub
\ba
\bs
B_{h400}^{E1} &=
\frac{3 I_1^2 \kappa^4 M^2}{40 \pi ^2 R}
+R^3 I_1 \kappa^2 \text{p}_0+\frac{I_1 \kappa^2 M \Pi _0}{4 \pi }  \\
&\phantom{=}+ \frac{9 I_1'^2 \kappa^4 M^2}{512 \pi^2 F m^5 R^6} \\
&\qquad \times \Bigg(
e^{-2 m R} \l(13+26 m R+15 m^2 R^2+2 m^3 R^3\r) - \left(13-11 m^2 R^2+\frac{20 m^3 R^3}{3}\right)
\Bigg)
\es
\\
B_{h400}^{E2} &= -\frac{I_1^2 \kappa^4 M^2}{32 \pi ^2}
\\
B_{h400}^{E3} &= \frac{9 I_1'^2 \kappa^4 M^2
\left( e^{-m R}(1+m R) + e^{m R} (-1+m R)\right)^2 }{1024 \pi^2 F m^5 R^6}
\\
B_{h400}^{E4} &= 2 m B_{h400}^{E3} \,.
\ea
\esub

\newpage

\subsection{Fourth order scalar}
\label{app coefficients fourth order scalar}

The coefficients for the fourth order scalar equation are, for the interior part \eqref{BETA equ phi4 equation int}
\bsub
\ba
A_{\phi 4}^{I1} &=
-\frac{9 I_1' \kappa^4 M^2 \left(4 F \left(I_1'^2+I_1'' I_1\right) m^2-I_1' I_2''' \kappa^2\right)}{1024 \pi^2 F^3 I_1 m^4 R^6}
-\frac{3 I_1' \kappa^2 p_0}{4 F}
+\frac{3 I_1' \kappa^2 M \Pi_0}{16 \pi F  R^3}
\\
A_{\phi 4}^{I2} &= \l( m^2 - \f{\k^2 I_2'''}{2 F' I_1} \r) A_{\phi 4}^{I3}
\\
A_{\phi 4}^{I3} &= \frac{9 F' I_1'^2 \kappa^4 M^2 e^{-2 m R} (1+m R)^2}{1024 \pi ^2 F^3 m^6 R^6}
\\
A_{\phi 4}^{I4} &= -\frac{9 I_1' \kappa^4 M^2  e^{-m R} (1+m R) \left(2 m^2 \left(F' I_1' I_1 + F I_1'^2 + F I_1 I_1''+2 F^2 I_1^2 m^2 R^2\right)-I_1' I_2''' \kappa^2 \right)}{1024 \pi^2 F^3 I_1 m^5 R^6}
\\
A_{\phi 4}^{I5} &= \frac{9 I_1'^2 \kappa^4 M^2 e^{-2 m R} (1+m R)^2 \left(-6 F' I_1 m^2+I_2''' \kappa^2 \right)}{4096 \pi^2 F^3 I_1 m^6 R^6}
\\
A_{\phi 4}^{I6} &= -m A_{\phi 4}^{I3}
\\
A_{\phi 4}^{I7} &= - \f12 A_{\phi 4}^{I3}
\\
A_{\phi 4}^{I8} &= - A_{\phi 4}^{I4}
\\
A_{\phi 4}^{I9} &= A_{\phi 4}^{I5}
\\
A_{\phi 4}^{I10} &= m A_{\phi 4}^{I3}
\\
A_{\phi 4}^{I11} &= - \f12 A_{\phi 4}^{I3}
\\
A_{\phi 4}^{I12} &= \frac{3 I_1' I_1 \kappa^4 M^2 e^{-m R} (1+m R) }{256 \pi^2 F m R^6}
\\
A_{\phi 4}^{I13} &= - A_{\phi 4}^{I12}
\ea
\esub
and for the exterior part \eqref{BETA equ phi4 equation ext}
\bsub
\ba
A_{\phi 4}^{E1} &= \frac{3 I_1' I_1 \kappa^4 M^2 (-m R \cosh (m R)+\sinh (m R))}{64 \pi^2 F m R^3}
\\
A_{\phi 4}^{E2} &= \frac{9 I_1'^2 \kappa^4 M^2 \left(-6 F' I_1 m^2+I_2''' \kappa^2 \right) (-m R \cosh (m R)+\sinh (m R))^2}{1024 \pi^2 F^3 I_1 m^6 R^6}
\\
A_{\phi 4}^{E3} &= -\frac{16 \pi^2 F'}{F I_1^2 m^3 \kappa^4 M^2} \l( A_{\phi 4}^{E1} \r)^2
\\
A_{\phi 4}^{E4} &= \frac{1}{2 m} A_{\phi 4}^{E3} \,.
\ea
\esub

For the interior solution \eqref{BETA equ phi4 solution int} they are
\bsub
\ba
\bs
B_{\phi 4}^{I1} &=
\frac{9 I_1' \kappa^4 M^2 \left(4 \text{F} \left(I_1'^2+I_1'' I_1\right) m^2-I_1' I_2''' \kappa^2\right)}{1024 \pi^2 \text{F}^3 I_1 m^6 R^6}
+\frac{3 I_1' \kappa^2 \text{p}_0}{4 \text{F} m^2}
-\frac{3 I_1' \kappa^2 M \Pi_0}{16 \pi \text{F} m^2 R^3}
\es
\\
B_{\phi 4}^{I2} &= \frac{9 F' I_1'^2 \kappa^4 M^2 e^{-2 m R} (1+m R)^2}{2048 \pi^2 F^3 m^6 R^6}
\\
\bs
B_{\phi 4}^{I3} &= \frac{9 I_1'^2 \kappa^4 M^2 e^{-2 m R} (1+m R)^2 \left(3 F' I_1 m^2-I_2''' \kappa^2\right) (2 \mathrm{Ei}(-m R)-\mathrm{Ei}(m R))}{8192 \pi^2 F^3 I_1 m^7 R^6} \\
&\phantom{=}+\frac{3 I_1 I_1' \kappa^4 M^2 \left( e^{-m R} ( 1+m R) + e^{m R} (-1+m R)\right) \mathrm{Ei}(-2 m R)}{256 \pi^2 F m^2 R^3} \\
&\phantom{=}+\frac{9 I_1'^2 \kappa^4 M^2 (-1+m R) \left(2+2 m R+e^{2 m R} (-1+m R)\right) \left(3 F' I_1 m^2-I_2''' \kappa^2\right) \mathrm{Ei}(-3 m R)}{8192 \pi^2 F^3 I_1 m^7 R^6} \\
&\phantom{=}+ \frac{e^{-3 m R} I_1' (1+m R) \kappa^2}{8192 \pi^2 F^3 I_1 m^7 R^6} \Big\{
2 \kappa^2 M^2
\Big[
18 \text{F} \left(I_1'^2+I_1'' I_1\right) m^2 \left(1+e^{2 m R} (5+2 m R)\right) \\
&\quad\quad\quad+2 \text{F}^2 I_1^2 m^2
\left(-3+6 m R (-1+2 m R)+e^{2 m R} \left(-3+2 m^2 R^2 (9+16 m R)\right)\right)\\
&\quad\quad\quad+9 I_1' \left(2 F' I_1 m^2 \left(1+2 e^{2 m R} m R\right)-I_2''' \left(1+e^{2 m R} (3+2 m R)\right) \kappa^2 \right)
\Big]\\
&\quad\quad+768 \pi F^2 I_1 m^4 R^3 e^{2 m R} \left(4 \pi \text{p}_0 R^3-M \Pi_0\right)
\Big\}
\es
\\
B_{\phi 4}^{I4} &= - \f12 B_{\phi 4}^{I2}
\\
B_{\phi 4}^{I5} &= - B_{\phi 4}^{I3}
\\
B_{\phi 4}^{I6} &= - \f12 B_{\phi 4}^{I2}
\\
B_{\phi 4}^{I7} &= \frac{3 I_1' \kappa^4 M^2 e^{-m R} (1+m R)}{2048 \pi^2 F^3 m^6 R^6 I_1}
\Bigg(
2 F^2 m^2 \left(-1+6 m^2 R^2\right) I_1^2+6 F' m^2 I_1 I_1' \\
&\qquad+6 F m^2 \left(I_1'^2+I_1 I_1''\right)-3 \kappa^2 I_1' I_2'''
\Bigg)
\\
B_{\phi 4}^{I8} &= -\frac{3 I_1 I_1' \kappa^4 M^2 e^{-m R} (1+m R)}{1024 \pi^2 F m^3 R^6}
\\
B_{\phi 4}^{I9} &= \f{2 m}{3} B_{\phi 4}^{I8}
\\
B_{\phi 4}^{I10} &= B_{\phi 4}^{I7}
\\
B_{\phi 4}^{I11} &= - B_{\phi 4}^{I8}
\\
B_{\phi 4}^{I12} &= \f{2 m}{3} B_{\phi 4}^{I8}
\\
B_{\phi 4}^{I13} &=\frac{9 I_1'^2 \kappa^4 M^2 e^{-2 m R} (1+m R)^2 \left(3 F' I_1 m^2-I_2''' \kappa^2\right)}{8192 \pi^2 F^3 I_1 m^7 R^6}
\\
B_{\phi 4}^{I14} &= 2 B_{\phi 4}^{I13}
\\
B_{\phi 4}^{I15} &= - B_{\phi 4}^{I13}
\\
B_{\phi 4}^{I16} &= - B_{\phi 4}^{I13}
\\
B_{\phi 4}^{I17} &= -2 B_{\phi 4}^{I13}
\\
B_{\phi 4}^{I18} &= B_{\phi 4}^{I13}
\ea
\esub
and for the exterior solution \eqref{BETA equ phi4 solution ext}
\bsub
\ba
\bs
B_{\phi 4}^{E1} &=
- B_{\phi 4}^{E4} \mathrm{Ei}(-2 m R) \\
&\phantom{=}+ \frac{9 I_1'^2 \kappa^4 M^2  \left(3 F' I_1 m^2 - I_2''' \kappa^2 \right)}{8192 \pi^2 F^3 I_1 m^7 R^6}
\Bigg(
\left(2-2 m^2 R^2-e^{2 m R} (-1+m R)^2\right) (-\mathrm{Ei}(-3 m R)+\mathrm{Ei}(-m R)) \\
&\qquad\qquad+e^{-2 m R} (1+m R)^2 (2 \mathrm{Ei}(-m R)-3 \mathrm{Ei}(m R)+\mathrm{Ei}(3 m R))
\Bigg)
\\
&\phantom{=}+\frac{e^{-3 m R} I_1' \kappa^2}{8192 \pi^2 F^3 I_1 m^7 R^6}\\
&\qquad\times\Bigg(
2 M^2 \kappa^2
\Bigg[
18 F \left(I_1'^2+I_1'' I_1\right) m^2 \left(1+m R+4 e^{2 m R} (1+m R)^2+e^{4 m R} (-5+3 m R)\right)\\
&\qquad\qquad\qquad+2 F^2 I_1^2 m^2 (1+m R) \big(-3-6 m R+12 m^2 R^2+64 e^{2 m R} m^3 R^3\\
&\qquad\qquad\qquad\qquad\qquad-3 e^{4 m R} \left(-1+2 m R+4 m^2 R^2\right)\big)\\
&\qquad\qquad\qquad+9 I_1' \big(2 F' I_1 m^2 \left(1-2 e^{4 m R}+m R+e^{2 m R} \left(1+5 m R+4 m^2 R^2\right)\right)\\
&\qquad\qquad\qquad-I_2''' \left(1+m R+e^{4 m R} (-3+m R)+e^{2 m R} \left(2+6 m R+4 m^2 R^2\right)\right) \kappa^2\big)
\Bigg] \\
&\qquad\qquad+ 768 \pi F^2 I_1 m^4 R^3 e^{2 m R} \left(1+m R+e^{2 m R} (-1+m R)\right) \left(4 \pi \text{p}_0  R^3-M \Pi_0\right)
\Bigg)
%
\es
\\
B_{\phi 4}^{E2} &= - \frac{16 \pi^2 F'}{F I_1^2 m^2 \kappa^4 M^2}
(B_{\phi 4}^{E4})^2
\\
B_{\phi 4}^{E3} &= \l( -\frac{3 m}{2}+\frac{I_2''' \kappa^2}{2 F' I_1 m} \r) B_{\phi 4}^{E2}
\\
B_{\phi 4}^{E4} &= -\frac{3 I_1' I_1 \kappa^4 M^2 \left( e^{-m R} (1+m R) + e^{m R} (-1+m R)\right)}{256 \pi^2 F m^2 R^3}
\\
B_{\phi 4}^{E5} &= - B_{\phi 4}^{E3} \,.
\\
B_{\phi 4}^{E6} &= - B_{\phi 4}^{E4} \,.
\ea
\esub

\newpage

\subsection{PPN Beta}
\label{app PPN Beta}

The coefficients in the expression~\eqref{BETA.equ: beta} for $\beta$ are given by

\bsub
\ba
\bs
C_{\b}^{E1} &= \f{12}{5 R}
+ \f{3}{4 (2\omega + 3) m^5 R^6}
\big\{
[39 + m^2 R^2(20mR - 33)]\\
&\qquad\qquad\qquad\qquad\qquad
-3 (1+mR)[13+mR(13+2mR)] e^{-2 m R} \big\}\,,
\es
\\
C_{\b}^{E2} &= 1\,,
\\
\bs
C_{\b}^{E3} &= \frac{1}{64 (2\omega + 3) m^7 R^6}
\Bigg\{
6 e^{-3 m R} (1+m R) \Big[6(m^2\sigma - \mu)\\
&\qquad+2 m^2 \left(-1-2 m R+4 m^2 R^2\right)\Big]\\
&\qquad+8 e^{-m R} (1+m R) \Big[32 m^5 R^3 + 18m^2\sigma(mR + 1) - 9\mu(2mR + 1)\Big]\\
&\qquad-12 e^{m R} \Big[3m^2\sigma(5 - 3mR) + 3\mu(mR - 3)\\
&\qquad+m^2 \left(-1+m R+6 m^2 R^2+4 m^3 R^3\right)\Big]\\
&\qquad-18\mu\Big[\left[2-2 m^2 R^2-e^{2 m R} (-1+m R)^2\right] [\mathrm{Ei}(-m R)-\mathrm{Ei}(-3m R)]\\
&\qquad+e^{-2 m R} (1+m R)^2 [2 \mathrm{Ei}(-m R)-3 \mathrm{Ei}(m R)+\mathrm{Ei}(3 m R)]\Big]\\
&\qquad+96 e^{-m R} m^5 R^3 \left[1+m R+e^{2 m R} (-1+m R)\right] \mathrm{Ei}(-2 m R)\Bigg\} \\
&\phantom{=} + \frac{6}{5 R} C_{\b}^{E4}\,,
\es
\\
C_{\b}^{E4} &= 6\f{m R \cosh(m R) - \sinh(m R)}{(2\omega + 3) m^3 R^3}\,,
\\
C_{\b}^{E5} &= -\l[\frac{(2\omega + 3)m}{8} + \frac{3}{5 R} \r] \l( C_{\b}^{E4} \r)^2\,,
\\
C_{\b}^{E6} &= \frac{(2\omega + 3)\sigma - 4}{16} \l( C_{\b}^{E4} \r)^2\,,
\\
C_{\b}^{E7} &= -\frac{\mu(2\omega + 3)}{32m} \l( C_{\b}^{E4} \r)^2\,,
\\
C_{\b}^{E8} &= -\f{m}{2} C_{\b}^{E4}\,,
\\
C_{\b}^{E9} &= -C_{\b}^{E7}\,,
\\
C_{\b}^{E10} &= -\frac{(2\omega + 3) m^2}{4} \l( C_{\b}^{E4} \r)^2\,,
\\
C_{\b}^{E11} &= \f{m}{2} C_{\b}^{E4}\,.
\ea
\esub

\newpage


\bibliography{biblio}

\end{document}